\documentclass[notitlepage,aps,prl,twocolumn,superscriptaddress,amsmath,showpacs,tightenlines,longbibliography,10pt]{revtex4-1}
\usepackage{amssymb}
\usepackage{amsmath}
\usepackage{dcolumn}
\usepackage{graphicx}
\usepackage{mathrsfs}
\usepackage{subfigure}
\usepackage{booktabs}
\usepackage{color}
\usepackage{docmute}
\usepackage{hyphenat}
\usepackage{CJKutf8}
\usepackage{multirow}
\usepackage{enumerate}

\newdimen\origiwspc
\newdimen\origiwstr

\newcommand{\average}[1]{\mbox{$\langle#1\rangle$}}

\makeatletter
\newcommand{\raisemath}[1]{\mathpalette{\raisem@th{#1}}}
\newcommand{\raisem@th}[3]{\raisebox{#1}{$#2#3$}}
\makeatother

\usepackage{url}
\usepackage[colorlinks]{hyperref}
\hypersetup{%
	plainpages=true,
	breaklinks=true,
	hypertexnames=false,
	pageanchor=true,
	bookmarksnumbered=true,
	colorlinks=true,
	linkcolor={blue},
	citecolor={red},
	urlcolor={blue},
	anchorcolor={black}
}

\begin{document}
	

\title{Beating the 3~dB Limit for Intracavity Squeezing \\
	 and Its Application to Nondemolition Qubit Readout}

\author{Wei Qin}
\affiliation{Theoretical Quantum Physics Laboratory, RIKEN Cluster
	for Pioneering Research, Wako-shi, Saitama 351-0198, Japan}

\author{Adam Miranowicz}
\affiliation{Theoretical Quantum Physics Laboratory, RIKEN Cluster
	for Pioneering Research, Wako-shi, Saitama 351-0198, Japan}
\affiliation{Institute of Spintronics and Quantum Information,
	Faculty of Physics, Adam Mickiewicz University, 61-614 Pozna\'{n}, Poland}

\author{Franco Nori}
\affiliation{Theoretical Quantum Physics Laboratory, RIKEN Cluster
	for Pioneering Research, Wako-shi, Saitama 351-0198, Japan}
\affiliation{RIKEN Center for Quantum Computing, Wako-shi, Saitama 351-0198, Japan}
\affiliation{Department of Physics, The University of Michigan,
	Ann Arbor, Michigan 48109-1040, USA}

\begin{abstract}
While the squeezing of a propagating field can, in principle, be made arbitrarily strong, the cavity-field squeezing is subject to the well-known $3$~dB limit, and thus has limited applications. Here, we propose the use of a fully quantum degenerate parametric amplifier (DPA) to beat this squeezing limit. Specifically, we show that by {\it simply} applying a two-tone driving to the signal mode, the pump mode can, {\it counterintuitively}, be driven by the photon loss of the signal mode into a squeezed steady state with, in principle, an {\it arbitrarily high} degree of squeezing. Furthermore, we demonstrate that this intracavity squeezing can increase the signal-to-noise ratio of longitudinal qubit readout {\it exponentially} with the degree of squeezing. Correspondingly, an improvement of the measurement error by {\it many orders of magnitude} can be achieved even for modest parameters. In stark contrast, using intracavity squeezing of the semiclassical DPA {\it cannot} practically
increase the signal-to-noise ratio and thus improve the measurement error. Our results extend the range of applications of DPAs and open up new opportunities for modern quantum technologies.
\end{abstract}

\date{\today}

\maketitle

\emph{Introduction.---}Squeezed states of light~\cite{drummond2004quantum} form a fundamental building block in modern quantum technologies ranging from quantum metrology~\cite{schnabel2017squeezed,lawrie2019quantum} to quantum information processing~\cite{braunstein2005quantum,weedbrook2012gaussian}. In particular, squeezing of a propagating field can, in principle, be made arbitrarily strong, due to destructive interference between the reflected input field and the transmitted cavity field; e.g., the squeezing of up to 15~dB has been experimentally achieved~\cite{vahlbruch2016detection}. Such a propagating-field squeezing has been widely used for, e.g., gravitational-wave
detection~\cite{rabl2010quantum,aasi2013enhanced,grote2013first}, mechanical cooling~\cite{asjad2016suppression,clark2017sideband}, nondemolition qubit readout~\cite{barzanjeh2014dispersive,didier2015heisenberg,didier2015fast,govia2017enhanced}, and even demonstrating quantum supremacy~\cite{zhong2020quantum,zhong2021phase}. However, these applications inherently suffer from transmission and injection losses, which are a major obstacle to using extremely fragile squeezed states. To address this problem, exploiting intracavity squeezing (i.e., squeezing of a cavity field) offers a promising route.

\begin{figure*}[t]
	\centering
	\includegraphics[width=17cm]{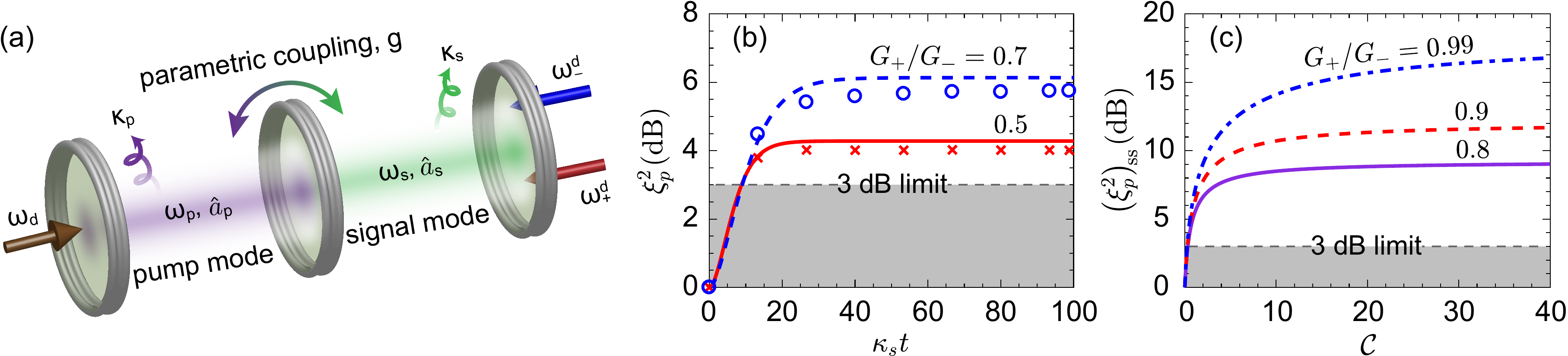}
	\caption{(a) Schematic of our proposal with a fully quantum DPA. We use two cavities to represent the pump mode $\hat{a}_{p}$ (frequency $\omega_{p}$, loss rate $\kappa_{p}$) and the signal mode $\hat{a}_{s}$ (frequency $\omega_{s}$, loss rate $\kappa_{s}$). The single-photon parametric coupling between them has a strength $g$. A driving tone at frequency $\omega_{d}$ is applied to the pump mode and, simultaneously, the signal mode is driven by the other two tones at frequencies $\omega_{\pm}^{\rm d}$. (b) Time evolution of the squeezing parameter $\xi^{2}_{p}$ for $G_{+}/G_{-}=0.5$ and $0.7$. We assumed that $\Delta_{s}=100g$, $\Delta_{p}=0.1\Delta_{s}$, $\Omega_{\rm 2pd}=0.05\Delta_{s}$, $\kappa_{s}=100\kappa_{p}=0.4g$, and $G_{-}=g_{0}$. Curves are the effective predictions, while symbols are the exact results. (c) Steady-state squeezing parameter $\left(\xi_{p}^{2}\right)_{\protect\raisemath{1.2pt}{\rm ss}}$ versus the cooperativity $\mathcal{C}$ for $\kappa_{s}=100\kappa_{p}$, and for $G_{+}/G_{-}=0.8$, $0.9$, and $0.99$. In (b) and (c), the gray shaded areas refer to the regime below the $3$~dB limit.}\label{fig_schematics}
\end{figure*}

To date, intracavity squeezing has been applied, e.g., to cool mechanical resonators~\cite{huang2009enhancement,asjad2019optomechanical,lau2020ground}, to enhance light-matter interactions~\cite{lu2015squeezed,zeytinouglu2017engineering,qin2018exponentially,leroux2018enhancing,ge2019trapped,li2020enhancing,burd2021quantum,tang2022quantum}, to improve high-precision measurements~\cite{zagoskin2008controlled,peano2015intracavity,eddins2019high}, and to generate nonclassical states~\cite{krippner1994transient,munro1995transient,groszkowski2020heisenberg,chen2021shortcuts,qin2021generating}. Despite such developments, the range and quality of applications of intracavity squeezing are still largely limited by the fact that quantum noise of a cavity field cannot be reduced below one-half
of the zero-point fluctuations in the steady state~\cite{drummond1981non,milburn1981production,collett1984squeezing}, i.e., the 3~dB limit. However, how to beat this limit has so far remained challenging, although for more complicated mechanical oscillators, the steady-state squeezing beyond 3~dB has been widely demonstrated both theoretically~\cite{kronwald2013arbitrarily,woolley2014two,kustura2022mechanical} and experimentally~\cite{szorkovszky2013strong,lei2016quantum}. The reason for the 3~dB limit of intracavity squeezing is the cavity photon loss, which is always present, destroys the essence of squeezing, i.e., two-photon correlations. In this manuscript, we show that, if such a photon loss is exploited as a resource, a strong steady-state intracavity squeezing can be achieved. 

In our approach, we consider a fully quantum DPA, where both pump and signal modes are quantized. We show that a strong photon loss of the signal mode can steer the pump mode into a squeezed steady state, with a noise level reduced far beyond $3$~dB. In this way, an {\it arbitrarily strong} steady-state squeezing of the pump mode can, in principle, be achieved. Note that optical experiments performed in the
	1990s (see, e.g.,~\cite{Paschotta1994,Ralph1995}) demonstrated
	bright squeezing of the pump mode (i.e., the second-harmonic mode)
	by driving the signal mode (i.e., the fundamental mode). However, it was achieved for output squeezing only, not for intracavity squeezing.

To beat the $3$~dB limit of intracavity squeezing, a theoretical approach, which requires a fast modulation of the coupling between the cavity and its environment, has been proposed~\cite{didier2014perfect}; and very recently, an experimental demonstration with three microwave modes coupled via a specific Josephson ring modulator was reported in Ref.~\cite{dassonneville2021dissipative}. In contrast, our approach relies only on common degenerate parametric amplification processes, and therefore is more compatible with current quantum technologies based on parametric amplification. More remarkably, we show that only a two-tone driving, if applied to the signal mode, can result in a strong steady-state squeezing for the pump mode. This is rather {\it counterintuitive}; indeed, common sense suggests that, as mentioned above, the steady-state intracavity squeezing of a DPA is usually limited to $3$~dB. We note that quantum intracavity noise reduction can also be	realized via squeezing of photon-number fluctuations, corresponding to the sub-Poissonian photon-number statistics or photon antibunching (see, e.g., the early predictions~\cite{Ritsch1990,Teich1988} and very recent	demonstrations of 3~dB squeezing like in~\cite{Carroll2021}).

Fast and high-fidelity nondemolition qubit readout is a prerequisite for quantum error correction~\cite{schindler2011experimental,kelly2015state} and fault-tolerant quantum computation~\cite{raussendorf2007fault,gambetta2017building}. Using squeezed light to improve such a readout is a long-standing goal~\cite{barzanjeh2014dispersive,didier2015fast,didier2015heisenberg,blais2021circuit}. However, the simplest strategy, i.e., dispersive qubit readout~\cite{blais2021circuit,krantz2019quantum}, induces a qubit-state-dependent rotation of squeezing, such that the amplified noise in the antisqueezed quadrature is introduced into the signal quadrature, ultimately limiting the improvement of the signal-to-noise ratio (SNR). Thus, related experimental demonstrations in this context have remained elusive. Until recently, an improvement, enabled by injecting squeezed light into a cavity, was realized~\cite{eddins2018stroboscopic} for longitudinal qubit readout~\cite{didier2015fast,blais2021circuit,gu2017microwave,touzard2019gated,ikonen2019qubit}, which can enable much
shorter measurement times than the dispersive readout. However, due to transmission and injection losses, more than half of the amount of squeezing is lost, and consequently the reported SNR is increased only by $\simeq25\%$.

Here, we propose to apply our strong intracavity squeezing to longitudinal qubit readout, thus avoiding transmission and injection losses. We demonstrate that the SNR can be increased  {\it exponentially}, and the measurement error is improved by {\it many orders of magnitude} for modest parameters. In sharp contrast, intracavity squeezing of the semiclassical DPA {\it cannot} significantly improve the SNR during a practically feasible measurement time, even though squeezing of the output field is very strong. Our main results are summarized in Table I in~\cite{supplement}.

\emph{Physical model.---}A fully quantum DPA, as shown in Fig.~\ref{fig_schematics}(a), consists of a pump mode $\hat{a}_{p}$ and a signal mode $\hat{a}_{s}$, which are coupled through a single-photon parametric coupling of strength $g$. We assume that the pump mode is driven by a tone of frequency $\omega_{d}$ and amplitude $\mathcal{E}_{d}$, and, additionally, the signal mode is subject to a two-tone driving of frequencies $\omega_{\pm}^{\rm d}$ and amplitudes $\mathcal{E}_{\pm}$. The corresponding Hamiltonian in a frame rotating at $\omega_{d}$ is  $\hat{H}=\hat{H}_{0}+\hat{H}_{\rm 2td}$, with
\begin{align}\label{eq:H_DPA_01}
\hat{H}_{0}=\;&\Delta_{p}\hat{a}_{p}^{\dagger}\hat{a}_{p}+\Delta_{s}\hat{a}_{s}^{\dagger}\hat{a}_{s}\nonumber\\
&+g\left(\hat{a}_{s}^{2}\hat{a}_{p}^{\dagger}+{\rm H.c.}\right)+\left(\mathcal{E}_{d}\hat{a}^{\dagger}_{p}+{\rm H.c.}\right),\\
\hat{H}_{\rm 2td}=\;&
\Omega_{\rm 2td}\left(t\right)\hat{a}_{s}^{\dagger}+{\rm H.c.},
\end{align} 
where $\Delta_{p}=\omega_{p}-\omega_{d}$, $\Delta_{s}=\omega_{s}-\omega_{d}/2$, $\Omega_{\rm 2td}\left(t\right)=\mathcal{E}_{-}\exp\left(-i\omega_{-}t\right)+\mathcal{E}_{+}\exp\left(-i\omega_{+}t\right)$. 
Here, $\omega_{p}$, $\omega_{s}$ are the resonance frequencies of the pump and signal modes, and $\omega_{\pm}=\omega_{\pm}^{\rm d}-\omega_{d}/2$. We describe photon losses with the Lindblad dissipator $\mathcal{L}\left(\hat{o}\right)\hat{\rho}=\hat{o}\hat{\rho} \hat{o}^{\dagger}-\frac{1}{2}\left(\hat{o}^{\dagger}\hat{o}\hat{\rho}+\hat{\rho} \hat{o}^{\dagger}\hat{o}\right)$, so that the system dynamics is determined by the master equation
$\dot{\hat{\rho}}=-i\left[\hat{H},\hat{\rho}\right]+\kappa_{p}\mathcal{L}\left(\hat{a}_{p}\right)\hat{\rho}+\kappa_{s}\mathcal{L}\left(\hat{a}_{s}\right)\hat{\rho}$,
where $\kappa_{p}$ and $\kappa_{s}$ are the photon-loss rates.
Upon introducing the displacement transformation $\hat{a}_{p}\rightarrow \hat{a}_{p}+\alpha_{p}^{\rm d}$, where $\alpha_{p}^{\rm d}=\mathcal{E}_{d}/\left(i\kappa_{p}/2-\Delta_{p}\right)$, the Hamiltonian $\hat{H}_{0}$ becomes $\hat{H}_{0}=\Delta_{p}\hat{a}_{p}^{\dagger}\hat{a}_{p}+\hat{H}_{\rm 2pd}+\hat{V}$. Here,
\begin{align}
\label{eq_semi_classical_parametric_amplifier}
\hat{H}_{\rm 2pd}=\;&\Delta_{s}\hat{a}_{s}^{\dagger}\hat{a}_{s}+\Omega_{\rm 2pd}\left(\hat{a}_{s}^{2}+{\rm H.c.}\right),\\
\label{eq_coupling_V}
\hat{V}=\;&g\left(\hat{a}_{s}^{2}\hat{a}_{p}^{\dagger}+{\rm H.c.}\right),
\end{align}
where $\Omega_{\rm 2pd}=g\alpha_{p}^{\rm d}$ can be viewed as the strength of a two-photon driving of the mode $\hat{a}_{s}$. We have assumed, for simplicity, that $\alpha_{p}^{\rm d}$ is real. 

Since the single-photon coupling $g$ is usually weak, the most studied regime of the DPA is for $\alpha_{p}^{\rm d}\gg1$. It is then standard to drop $\hat{V}$, leaving only $\hat{H}_{\rm 2pd}$. In this case, the pump mode is treated as a classical field, and the DPA is referred to as semiclassical. For such a semiclassical DPA, the signal mode cannot be squeezed above 3~dB, even with nonlinear corrections arising from the coupling $\hat{V}$~\cite{milburn1981production,chaturvedi2002limits,supplement}. The reason for this moderate squeezing is the photon loss of the signal mode. That is, the leakage of single photons of some correlated photon pairs injected by the two-photon driving $\Omega_{\rm 2pd}$ causes a partial loss of two-photon correlations,  and thus of intracavity squeezing. However, as demonstrated below, the photon loss of the signal mode, when turned from a noise source into a resource via reservoir engineering, can steer a quantized pump mode into a squeezed steady state. More importantly, this photon loss can strongly suppress the detrimental effect of the photon loss of the pump mode on squeezing, ultimately leading to a strong steady-state intracavity squeezing.

\emph{Squeezing far beyond 3~dB.---}Recently, it has been shown experimentally that the available single-photon coupling $g$ can range from tens of kHz to tens of MHz~\cite{leghtas2015confining,touzard2018coherent,lescanne2020exponential,chang2020observation,vrajitoarea2020quantum,guo2016second,bruch2019chip,wang2021efficient,nehra2022few}. These advances allow one to consider the effect of the coupling $\hat{V}$, e.g., two-photon loss~\cite{leghtas2015confining,touzard2018coherent,lescanne2020exponential,everitt2014engineering,mirrahimi2014dynamically,sun2019schrodinger,sun2019discrete}. We here focus on the case of $\Delta_{s}\neq0$, and introduce a signal Bogoliubov mode, $\hat{\beta}_{s}=\hat{a}_{s}\cosh\left(r_{s}\right)+\hat{a}_{s}^{\dagger}\sinh\left(r_{s}\right)$, with $\tanh\left(2r_{s}\right)=2\Omega_{\rm 2pd}/\Delta_{s}$.
The Hamiltonian $\hat{H}_{\rm 2pd}$ is then diagonalized, yielding $\hat{H}_{\rm 2pd}=\Lambda_{s}\hat{\beta}_{s}^{\dagger}\hat{\beta}_{s}$, where $\Lambda_{s}=\sqrt{\Delta_{s}^{2}-4\Omega_{\rm 2pd}^{2}}$. Likewise, the coupling $\hat{V}$ and the two-tone driving $\hat{H}_{\rm 2td}$ become
\begin{align}\label{eq_H2td_V}
\hat{V}=\;&g_{0}\hat{\beta}_{s}^{\dagger}\hat{\beta}_{s}\left(\hat{a}_{p}+\hat{a}_{p}^{\dagger}\right)+\hat{R}_{1}+\hat{R}_{1}^{\dagger},\\
\hat{H}_{\rm 2td}=\;&\Omega_{\rm 2td}\left(t\right)\cosh\left(r_{s}\right)\hat{\beta}_{s}^{\dagger}+\hat{R}_{2}+{\rm H.c.},
\end{align}
where $\hat{R}_{1}=g\left[\cosh^{2}\left(r_{s}\right)\hat{\beta}_{s}^{2}+\sinh^{2}\left(r_{s}\right)\hat{\beta}_{s}^{\dagger2}\right]\hat{a}_{p}^{\dagger}$, $\hat{R}_{2}=-\Omega_{\rm 2td}\left(t\right)\sinh\left(r_{s}\right)\hat{\beta}_{s}$, and $g_{0}=-g\sinh\left(2r_{s}\right)$. We further assume the limit $\left\{g, \Omega_{\rm 2pd},\Delta_{p}\right\}\ll\Delta_{s}$, such that $r_{s}\ll1$, and both $\hat{R}_{1}$ and $\hat{R}_{2}$ can be dropped as high-frequency components (see~\cite{supplement}), yielding
\begin{align}
\label{eq_V0}
\hat{V}\simeq\;&g_{0}\hat{\beta}_{s}^{\dagger}\hat{\beta}_{s}\left(\hat{a}_{p}+\hat{a}_{p}^{\dagger}\right),\\
\label{eq_two_tone_driving_betas}
\hat{H}_{\rm 2td}\simeq\;&\cosh\left(r_{s}\right)
\Omega_{\rm 2td}\left(t\right)\hat{\beta}_{s}^{\dagger}+{\rm H.c.}
\end{align}

Equations~(\ref{eq_V0}), (\ref{eq_two_tone_driving_betas}) are reminiscent of the two-tone driven radiation-pressure interaction in cavity optomechanics~\cite{aspelmeyer2014cavity}. With such an interaction, the cavity photon loss can stabilize a strong squeezing of mechanical motion~\cite{kronwald2013arbitrarily,woolley2014two,wollman2015quantum,pirkkalainen2015squeezing,lei2016quantum,ockeloen2018stabilized}. Here, we harness a similar mechanism, and assume that $\omega_{\pm}=\Lambda_{s}\pm\Delta_{p}$, so that the mode $\hat{\beta}_{s}$ is coupled to a pump Bogoliubov mode,
$\hat{\beta}_{p}=\hat{a}_{p}\cosh\left(r_{p}\right)+\hat{a}_{p}^{\dagger}\sinh\left(r_{p}\right)$, through the effective Hamiltonian~\cite{supplement},
\begin{align}\label{eq_squeezing_Hamiltonian}
\hat{H}_{\rm eff}=\mathcal{G}\left(\hat{\beta}_{p}\hat{\beta}_{s}^{\dagger}+\hat{\beta}_{p}^{\dagger}\hat{\beta}_{s}\right).
\end{align}
Here, $\tanh(r_{p})=G_{+}/G_{-}$ and $\mathcal{G}=\sqrt{G_{-}^{2}-G_{+}^{2}}$. We have defined $G_{\pm}= g_{0}\alpha^{\pm}_{s}$, where  $\alpha_{s}^{\pm}$ (given in~\cite{supplement}) are the field amplitudes of the mode $\hat{\beta}_{s}$ induced by the two-tone driving $\Omega_{\rm 2td}$, and for simplicity both have been assumed to be real. 

Furthermore, we have $\mathcal{L}\left(\hat{a}_{s}\right)\hat{\rho}\simeq\mathcal{L}\left(\hat{\beta}_{s}\right)\hat{\rho}$ for $r_{s}\ll1$, and the system dynamics can thus be described with the effective master equation
\begin{equation}\label{eq_effective_master_equation}
\dot{\hat{\rho}}=-i\left[\hat{H}_{\rm eff},\hat{\rho}\right]+\kappa_{p}\mathcal{L}\left(\hat{a}_{p}\right)\hat{\rho}+\kappa_{s}\mathcal{L}(\hat{\beta}_{s})\hat{\rho}.
\end{equation}
It is seen that for a large $\kappa_{s}$, the photon loss of the mode $\hat{\beta}_{s}$ can cool the mode $\hat{\beta}_{p}$ into the ground state, corresponding to the squeezed vacuum state of the mode $\hat{a}_{p}$, which can theoretically have an arbitrary degree of squeezing. Such a squeezed steady state is unique, and can be reached from any state of the mode $\hat{a}_{p}$. The reason is that any state of the mode $\hat{a}_{p}$ can be expressed in terms of the ground and excited states of the mode $\hat{\beta}_{p}$, but of these, all the excited states are depopulated by the photon loss of the mode $\hat{\beta}_{s}$ in the steady state. This initial-state independence enables the detrimental effect of the photon loss of the mode $\hat{a}_{p}$ on squeezing to be strongly suppressed as long as $\kappa_{s}\gg\kappa_{p}$~(see~\cite{supplement} for more details), consequently, forming a strong steady-state squeezing for the mode $\hat{a}_{p}$. During the formation of this squeezing, any odd photon-number state of the mode $\hat{a}_{p}$ is reached by two different transitions, which are induced by the two-tone driving $\Omega_{\rm 2td}$. Achieving a desired steady-state squeezing, i.e., a superposition of only even photon-number states, requires destructive interference between these two transitions to cancel out the population of all the odd photon-number states.

\begin{figure*}[t]
	\centering
	\includegraphics[width=17.5cm]{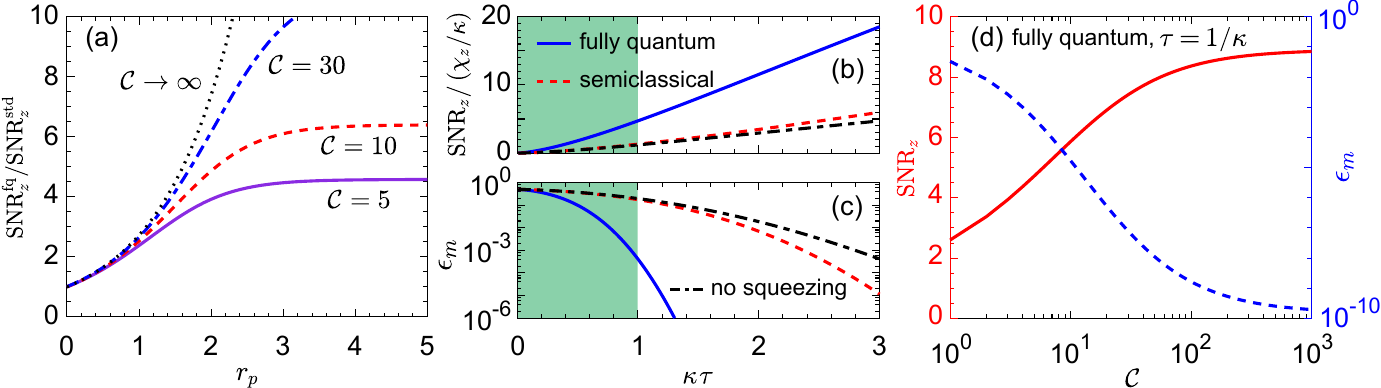}
	\caption{(a) SNR improvement, i.e., ${\rm SNR}_{z}^{\rm fq}/{\rm SNR}_{z}^{\rm std}$, versus the degree $r_{p}$ of intracavity squeezing for different values of the DPA cooperativity: $\mathcal{C}=5$, $10$, $30$, and $\infty$. An exponential improvement can be obtained for $\mathcal{C}\gg\exp\left(2r_{p}\right)/4$. (b) SNR and (c) measurement error versus the measurement time. The solid and dashed curves correspond to the longitudinal readout using intracavity squeezing of the fully quantum ($r_{p}=2$, $\mathcal{C}=5$) and semiclassical ($r_{\rm out}^{\rm sc}=2$)  DPAs, respectively, while the dash-dotted curves are results of the standard longitudinal readout with no squeezing. The green shaded area represents the experimentally most interesting regime. In (b) and (c), all the parameters are the same except that $\chi_{z}=\kappa$ in (c). (d) SNR (left axis) and measurement error (right axis) versus $\mathcal{C}$ in the fully quantum-DPA case for $\tau=1/\kappa$. Other parameters are the same as in (c).}\label{fig_measurement_error}
\end{figure*}

To quantify the degree of squeezing, we use the
squeezing parameter~\cite{ma2011quantum},
\begin{equation}\label{eq_squeezing_parameter}
\xi^{2}_{p}=1+2\left(\average{\hat{a}_{p}^{\dagger}\hat{a}_{p}}-\left|\average{\hat{a}_{p}\hat{a}_{p}}\right|\right).
\end{equation}
Its time evolution is plotted in Fig.~\ref{fig_schematics}(b). Specifically, we compare the effective and exact results, and show an excellent agreement between them. Therefore, the effective master equation in Eq.~(\ref{eq_effective_master_equation}) can be used to predict some larger squeezing by deriving the steady-state squeezing parameter,
\begin{align}\label{eq_steady_state_squeezing_parameter}
\left(\xi_{p}^{2}\right)_{\raisemath{1.2pt}{\rm ss}}=\frac{1+4\mathcal{C}\exp\left(-2r_{p}\right)}{1+4\mathcal{C}},
\end{align}
where $\mathcal{C}=\mathcal{G}^{2}/\left(\kappa_{s}\kappa_{p}\right)$ is the cooperativity of the DPA.
In Fig.~\ref{fig_schematics}(c), $\left(\xi_{p}^{2}\right)_{\raisemath{1.2pt}{\rm ss}}$ is plotted versus $\mathcal{C}$. For realistic parameters of $\kappa_{s}=100\kappa_{p}$, we find that a modest ratio $G_{+}/G_{-}$ can keep $\left(\xi_{p}^{2}\right)_{\raisemath{1.2pt}{\rm ss}}$ above $3$~dB even for $\mathcal{C}\simeq0.4$. Moreover, $\left(\xi_{p}^{2}\right)_{\raisemath{1.2pt}{\rm ss}}$ increases as $\mathcal{C}$, and ultimately reaches its maximum value,
\begin{equation}
\left(\xi_{p}^{2}\right)_{\raisemath{1.2pt}{\rm ss}}^{\raisemath{-1.2pt}{\rm max}}=\exp\left(-2r_{p}\right)=\frac{1-G_{+}/G_{-}}{1+G_{+}/G_{-}}.
\end{equation}
For example, with $G_{+}/G_{-}=0.99$, we predict a maximum squeezing of $\left(\xi_{p}^{2}\right)_{\raisemath{1.2pt}{\rm ss}}^{\raisemath{-1.2pt}{\rm max}}\simeq23$~dB. Thus by increasing the ratio $G_{+}/G_{-}$ to $\lesssim1$,
we can, in principle, make intracavity squeezing arbitrarily strong. This is a counterintuitive result from the usually accepted point of view: the steady-state intracavity squeezing of a DPA is fundamentally limited to $3$~dB. 

\emph{Enhanced longitudinal qubit readout.---}As an application, we below show that our intracavity squeezing in a fully quantum DPA can {\it exponentially} improve the SNR of longitudinal qubit readout. In~\cite{supplement}, we also analyze the longitudinal readout using intracavity squeezing of a semiclassical DPA. However, we demonstrate that this semiclassical-DPA intracavity squeezing {\it cannot} enable a practically useful increase in the SNR, even with a strong squeezing of the output field.

To begin, we consider the Hamiltonian
\begin{equation}\label{eq_Hamiltonian_longitudinal_readout}
\hat{H}_{z}^{\rm fq}=\hat{H}_{\rm eff}+\chi_{z}\hat{\sigma}_{z}\left(\hat{a}_{p}e^{-i\phi_{z}}
+\hat{a}^{\dagger}_{p}e^{i\phi_{z}}\right),
\end{equation}
where $\hat{\sigma}_{z}$ is the Pauli matrix of the qubit.
The first term is used to generate intracavity squeezing, while the second term accounts for the longitudinal qubit-field coupling of strength $\chi_{z}$ and phase $\phi_{z}$. 
Possible experimental implementations of $\hat{H}_{z}^{\rm fq}$ are discussed in~\cite{supplement}. Since the photon loss of the mode $\hat{\beta}_{s}$ is strong, we adiabatically eliminate the mode $\hat{\beta}_{s}$ to obtain the following equation of motion for the mode $\hat{a}_{p}$, 
\begin{equation}\label{eq_adiabatic_a_p}
\dot{\hat{a}}_{p}=-ie^{i\phi_{z}}\chi_{z}\hat{\sigma}_{z}-\frac{\kappa}{2}\hat{a}_{p}-\sqrt{\kappa}\hat{\mathcal{A}}_{\rm in}\left(t\right),
\end{equation}
where $\kappa=\kappa_{p}^{\rm ad}+\kappa_{p}$ is the overall photon loss rate. Here, $\kappa_{p}^{\rm ad}=4\mathcal{G}^{2}/\kappa_{s}$ is the rate of the adiabatic photon loss. Moreover, we have defined the overall input noise as
$\hat{\mathcal{A}}_{\rm in}\left(t\right)=\left[\sqrt{\kappa_{p}^{\rm ad}}\hat{a}_{p,{\rm in}}^{\rm ad}\left(t\right)+\sqrt{\kappa_{p}}\hat{a}_{p,{\rm in}}\left(t\right)\right]/\sqrt{\kappa}$.
It involves two uncorrelated noise operators, $\hat{a}_{p,{\rm in}}^{\rm ad}\left(t\right)$ and $\hat{a}_{p,{\rm in}}\left(t\right)$. The former represents the adiabatic noise arising from the photon loss of the mode $\hat{\beta}_{s}$, and is given by
$i\hat{a}_{p,{\rm in}}^{\rm ad}\left(t\right)=\hat{\beta}_{s,{\rm in}}\left(t\right)\cosh\left(r_{p}\right)+\hat{\beta}_{s,{\rm in}}^{\dagger}\left(t\right)\sinh\left(r_{p}\right)$,
where $\hat{\beta}_{s,{\rm in}}\left(t\right)$ is the noise operator of the mode $\hat{\beta}_{s}$. As seen in Eq.~(\ref{eq_effective_master_equation}), $\hat{\beta}_{s,{\rm in}}\left(t\right)$ can be considered as the vacuum noise, and therefore $\hat{a}_{p,{\rm in}}^{\rm ad}\left(t\right)$ corresponds to the squeezed vacuum noise of the mode $\hat{a}_{p}$. Moreover, the operator $\hat{a}_{p,{\rm in}}\left(t\right)$ represents the vacuum noise inducing the natural photon loss of the mode $\hat{a}_{p}$. Note that $\hat{a}_{p}$ in Eq.~(\ref{eq_adiabatic_a_p}) is a field operator displaced by an amount $\alpha_{p}^{\rm d}$, but the side effect of this displacement on the qubit readout is negligible as a high-frequency effect~\cite{supplement}.

The longitudinal coupling maps the qubit state onto the output quadrature, $\hat{\mathcal{Z}}_{\rm out}\left(t\right)=\hat{\mathcal{A}}_{\rm out}\left(t\right)e^{-i\phi_{h}}+\hat{\mathcal{A}}_{\rm out}^{\dagger}\left(t\right)e^{i\phi_{h}}$,
which is measured by a homodyne setup with a detection angle $\phi_{h}$. Here, 
$\hat{\mathcal{A}}_{\rm out}\left(t\right)=\hat{\mathcal{A}}_{\rm in}\left(t\right)+\sqrt{\kappa}\hat{a}_{p}\left(t\right)$ is the overall output field. An essential parameter quantifying the homodyne detection is the SNR, which is evaluated using the operator $\hat{M}=\sqrt{\kappa}\int_{0}^{\tau}dt\,\hat{\mathcal{Z}}_{\rm out}\left(t\right)$, with $\tau$ the measurement time, and is defined as 
\begin{equation}
{\rm SNR}=\left|\average{\hat{M}}_{\uparrow}-\average{\hat{M}}_{\downarrow}\right|\left(\average{\hat{M}_{N}^{2}}_{\uparrow}+\average{\hat{M}_{N}^{2}}_{\downarrow}\right)^{-1/2},
\end{equation}
where $\hat{M}_{N}=\hat{M}-\average{\hat{M}}$ characterizes the measurement noise, and $\{\uparrow, \downarrow\}$ refers to the qubit state. The SNR of the readout using our fully quantum-DPA intracavity squeezing is then given by
\begin{equation}\label{eq_SNR_enhancement}
{\rm SNR}_{z}^{\rm fq}=\sqrt{\frac{1+4\mathcal{C}}{1+4\mathcal{C}\exp\left(-2r_{p}\right)}}{\rm SNR}_{z}^{\rm std},
\end{equation}
where ${\rm SNR}_{z}^{\rm std}=8\chi_{z}\tau\left[1-2\left(1-e^{-\kappa\tau/2}\right)/\kappa\tau\right]/\sqrt{2\kappa\tau}$ refers to the SNR of the standard longitudinal readout with no squeezing.
Equation~(\ref{eq_SNR_enhancement}) shows a distinct improvement in the SNR, as in Fig.~\ref{fig_measurement_error}(a). Such an improvement increases as the cooperativity $\mathcal{C}$, which can, in principle, be made arbitrarily large.
Furthermore, as long as $\mathcal{C}\gg\exp\left(2r_{p}\right)/4$, we have
\begin{align}\label{eq_exponential_enhancement}
{\rm SNR}_{z}^{\rm fq}\simeq\exp\left(r_{p}\right){\rm SNR}_{z}^{\rm std},
\end{align}
an exponential improvement in the SNR.

More importantly, the SNR improvement in Eqs.~(\ref{eq_SNR_enhancement}), (\ref{eq_exponential_enhancement}) holds for {\it any} measurement time. The reason is that the degree of squeezing of the measurement noise equals the degree of intracavity squeezing, i.e., $\average{\hat{M}_{N}^{2}}/\kappa\tau=\left(\xi_{p}^{2}\right)_{\raisemath{1.2pt}{\rm ss}}$, and is independent of the measurement time. This is in stark contrast to the case of using the semiclassical-DPA intracavity squeezing, where, as discussed in~\cite{supplement}, the degree of squeezing of the measurement noise increases from the initial value zero, as the measurement time increases, and consequently a large increase in the SNR needs an extremely long measurement time. Assuming realistic parameters of $r_{p}=2$ ($\simeq17$~dB) and $\mathcal{C}=5$, our approach gives an approximately fourfold improvement for any measurement time, as illustrated in Fig.~\ref{fig_measurement_error}(a). However, when using the semiclassical-DPA intracavity squeezing, there is almost no improvement for the short-time measurement of most interest in experiments, even though  the output-field squeezing, characterized by the parameter $r_{\rm out}^{\rm sc}=\ln\left[(\kappa_{s}+4\Omega_{\rm 2pd})/(\kappa_{s}-4\Omega_{\rm 2pd})\right]$, is strong~\cite{supplement}. 

In Figs.~\ref{fig_measurement_error}(b),~\ref{fig_measurement_error}(c), we plot the SNR and the measurement error, $\epsilon_{m}=1-\mathcal{F}_{m}$, for the longitudinal readout using the fully quantum- and semiclassical-DPA intracavity squeezing, and also for the standard longitudinal readout with no squeezing. Here, $\mathcal{F}_{m}=\frac{1}{2}\left[1+{\rm erf}\left({\rm SNR}/2\right)\right]$ is the measurement fidelity.
Choosing $r_{p}=2$, and $\chi_{z}=\kappa=2\pi\times3$~MHz for our approach, a short measurement time of $\tau=1/\kappa\simeq53$~ns gives ${\rm SNR}^{\rm fq}_{z}\simeq4.7$ for $\mathcal{C}=5$. This corresponds to a measurement error of $\epsilon_{m}\simeq4.4\times10^{-4}$. When $\mathcal{C}$ increases, as in Fig.~\ref{fig_measurement_error}(d), ${\rm SNR}^{\rm fq}_{z}$ can further increase to a maximum of $\simeq8.9$, and the measurement error rapidly decreases, reaching a minimum of $\simeq1.5\times10^{-10}$. However, at the same measurement time, both the standard longitudinal readout with no squeezing and the case of using the semiclassical-DPA intracavity squeezing enable a much lower SNR, i.e., ${\rm SNR}^{\rm std}_{z}\simeq{\rm SNR}^{\rm sc}_{z}\simeq1.1$, and, correspondingly, a measurement error of $\simeq0.22$, which is many orders of magnitude larger. 

\emph{Conclusions.---}We have introduced a method of how to exploit a fully quantum DPA to beat the $3$~dB limit of intracavity squeezing. We have demonstrated that an {\it arbitrary} steady-state squeezing can, in principle, be achieved for the pump mode, by simply applying a two-tone driving to the signal mode. This counterintuitive intracavity squeezing can {\it exponentially} increase the SNR of longitudinal qubit readout, and improve the measurement error by {\it many orders of magnitude}. In contrast, the semiclassical-DPA intracavity squeezing {\it cannot} enable a useful increase in the SNR, due to the impractical requirement of a long measurement time. Our proposal is valid for both microwave and optical cavities, but we believe that it is easier to implement it with microwaves in quantum circuits. The resulting intracavity squeezing is equivalent to an externally generated and injected squeezing but without transmission and injection losses. Thus, this intracavity squeezing, as a powerful alternative to that external squeezing, could find many quantum applications in addition to the qubit readout, and further excite more interest to exploit the potential of DPAs for modern quantum technologies.

\begin{acknowledgments}
W.Q. was supported in part by the Incentive Research Project of RIKEN. A.M. was supported by the Polish National Science Centre (NCN) under the
Maestro Grant No. DEC-2019/34/A/ST2/00081.
F.N. is supported in part by: 
Nippon Telegraph and Telephone Corporation (NTT) Research, 
the Japan Science and Technology Agency (JST) [via 
the Quantum Leap Flagship Program (Q-LEAP), 
and the Moonshot R\&D Grant Number JPMJMS2061], 
the Japan Society for the Promotion of Science (JSPS) 
[via the Grants-in-Aid for Scientific Research (KAKENHI) Grant No. JP20H00134],
the Army Research Office (ARO) (Grant No. W911NF-18-1-0358),
the Asian Office of Aerospace Research and Development (AOARD) (via Grant No. FA2386-20-1-4069), and 
the Foundational Questions Institute Fund (FQXi) via Grant No. FQXi-IAF19-06.
\end{acknowledgments}


%

\end{document}


\begin{CJK*}{UTF8}{gbsn}

\title{Supplemental Material to:\\
	``Beating the 3~dB Limit for Intracavity Squeezing \\
	and Its Application to Nondemolition Qubit Readout"}

\author{Wei Qin}
\affiliation{Theoretical Quantum Physics Laboratory, RIKEN Cluster
	for Pioneering Research, Wako-shi, Saitama 351-0198, Japan}

\author{Adam Miranowicz}
\affiliation{Theoretical Quantum Physics Laboratory, RIKEN Cluster
	for Pioneering Research, Wako-shi, Saitama 351-0198, Japan}
\affiliation{Institute of Spintronics and Quantum Information,
	Faculty of Physics, Adam Mickiewicz University, 61-614 Pozna\'{n}, Poland}

\author{Franco Nori}
\affiliation{Theoretical Quantum Physics Laboratory, RIKEN Cluster
	for Pioneering Research, Wako-shi, Saitama 351-0198, Japan}
\affiliation{RIKEN Center for Quantum Computing, Wako-shi, Saitama 351-0198, Japan}
\affiliation{Department of Physics, The University of Michigan,
	Ann Arbor, Michigan 48109-1040, USA}

\makeatletter
\def\@hangfrom@section#1#2#3{\@hangfrom{#1#2#3}}
\makeatother

\maketitle

\section*{I\lowercase{ntroduction}}
\begin{quote}
Here, we first summarize our main results in this work. Next, we recall the $3$~dB limit of intracavity squeezing of a semiclassical degenerate parametric amplifier (DPA) to explain the reason for that limit, and accordingly to give the motivation of our method. Subsequently, we present a detailed derivation of the effective Hamiltonian of generating a strong steady-state intracavity squeezing,
and discuss its physical interpretation in the laboratory frame. Then, we demonstrate the physical mechanism of beating the 3~dB squeezing limit with our method. Furthermore, we analyze the effects of intracavity squeezing of the semiclassical and fully quantum DPAs on longitudinal qubit readout. Finally, a possible implementation of the intracavity-squeezing enhanced longitudinal qubit readout is discussed.

Click \href{https://www.dropbox.com/s/btx6ypoi65o31lp/talk_Caltech.pdf?dl=0}{here} to see a PDF file with slides about the present work, which is placed in Dropbox. 

The supplemental material also contains a pedagogical presentation of this work (see~\href{https://dml.riken.jp/wp-content/uploads/May2022_Beatingthe3dB_WeiQ_part1.mp4}{video1} and \href{https://dml.riken.jp/wp-content/uploads/May2022_Beatingthe3dB_WeiQ_part2.mp4}{video2}), available for interested readers. 

\end{quote}

\section{Summary of our main results}
\label{comparison}
Before discussing more technical details, we first summarize the main results of our work in this section. These main results, including some numerical values, are summarized in Table~\ref{stab:table} by comparing intracavity squeezing of the semiclassical and fully quantum DPAs, and their effects on longitudinal qubit readout. More details are described as follows:

\begin{table*}[b]
 	\centering
	\caption{Comparison between intracavity squeezing of the semiclassical and fully quantum DPAs, and between their effects on longitudinal qubit readout. Here, $\mathcal{C}$ is the cooperativity of the fully quantum DPA, SNR stands for the signal-to-noise ratio, ${\rm SNR}^{\rm std}_{z}$ is the SNR of the standard longitudinal readout with no squeezing, $\epsilon_{m}$ is the measurement error, $\tau$ is the measurement time, $r_{p}$ is the degree of intracavity squeezing of the fully quantum DPA, and $r_{\rm out}^{\rm sc}$ is the degree of squeezing of the output field of the semiclassical DPA.}
	\begin{threeparttable}
		\vspace*{0.2cm}
		\setlength{\tabcolsep}{1.1mm}\renewcommand{\arraystretch}{2.5}{
			\begin{tabular}{|c|c|c|c|c|c|c|c|}
				\hline 
				\hline 
				\multirow{2}{*}{DPA type}&
				\multirow{2}{*}{\makecell[c]{Steady-state \\ intracavity squeezing}}&
				\multicolumn{6}{c|}{Longitudinal qubit readout}\\
				\cline{3-8}
				& & $\mathcal{C}$ & ${\rm SNR}/{\rm SNR}_{z}^{\rm std}$ & SNR\tnote{$\S$} & $\epsilon_{m}$ & $\tau$~(ns) & \makecell[c]{Degree of squeezing of \\ the measurement noise} \\
				\hline
				semiclassical & limited to $3$~dB  & --- & $\simeq1$\tnote{$\dag$} &1.1& $\simeq0.22$ & \multirow{3}{*}{$\simeq53$} & $\tau$-dependent\\
				\cline{1-6}	\cline{8-8}		
				\multirow{2}{*}{fully quantum} & \multirow{2}{*}{arbitrarily strong} & any &  $\sqrt{\frac{4\mathcal{C}+1}{4\mathcal{C}\exp\left(-2r_{p}\right)+1}}$ & 4.7 & $\simeq4.4\times10^{-4}$ & & \multirow{2}{*}{$\tau$-independent} \\
				\cline{3-6}
				& & $\gg\frac{1}{4}\exp\left(2r_{p}\right)$ & $\simeq\exp\left(r_{p}\right)$ & 8.9 & $\simeq1.5\times10^{-10}$ & &\\
				\hline	
		\end{tabular}}\label{stab:table}
		\begin{tablenotes}
			\item[$\dag$] This case of effectively no improvement is observed even for a large $r_{\rm out}^{\rm sc}$;
			\item[$\S$] To calculate the SNR values, we assume $r_{\rm out}^{\rm sc}=r_{p}=2$ ($\simeq17$~dB), and a modest cooperativity of $\mathcal{C}=5$. 
		\end{tablenotes}
	\end{threeparttable}
\end{table*}

\begin{enumerate}[(1)]

\item While intracavity squeezing of a semiclassical DPA is limited, as is well known, to $3$~dB in the steady state, we show that our approach can, in principle, lead to an arbitrarily strong steady-state intracavity squeezing by exploiting the pump mode of a fully quantum DPA. 

The reason for this sharp contrast is that the photon loss of the signal mode, which is the origin of the $3$~dB squeezing limit of the semiclassical DPA, is now engineered as a resource in our approach rather than a noise source.

Our present work is, to our knowledge, the first demonstration of the possibility of exploiting a fully quantum DPA to beat the 3~dB limit for intracavity squeezing.

\item Furthermore, our strong fully-quantum-DPA intracavity squeezing, which is equivalent to an externally generated and injected squeezed reservoir, can be applied to longitudinal qubit readout. For comparison, we also analyze the effect of the semiclassical-DPA intracavity squeezing on such a readout. We find that these two types of intracavity squeezing exhibit strikingly different performances. 

In the semiclassical-DPA case, we find that the degree of squeezing of the measurement noise strongly depends on the measurement time, i.e., it increases from zero as the measurement time increases. This means that a large squeezing of the measurement noise, corresponding to a high readout SNR, needs a long measurement time. Thus, the semiclassical-DPA intracavity squeezing cannot enable a significant increase in the SNR for the short-time measurement of practical interest, even with a strong squeezing of the output field. However, this is not the case when our fully-quantum-DPA intracavity squeezing is used. 

We demonstrate that in the fully-quantum-DPA case, the degree of squeezing of the measurement noise is always equal to the degree of intracavity squeezing, and is independent of the measurement time. This enables an exponential increase in the SNR for any measurement time, and leads to an extremely low measurement error, which can be made many orders of magnitude smaller than those obtained in both the semiclassical-DPA case and the standard longitudinal readout with no squeezing. 

\end{enumerate}

Having summarized the main results presented in our work, we show more technical details in the following sections. 

\section{Limited steady-state intracavity squeezing of the semiclassical degenerate parametric amplifier}
\label{appendix_Limited steady-state intracavity squeezing of the signal mode of the semiclassical parametric amplifier}

In this section, we recall the $3$~dB limit of the semiclassical-DPA intracavity squeezing to explain the reason for such a limit, i.e., the photon loss of the signal mode and, accordingly, to give the motivation of our method. We begin with the standard master equation,
\begin{align}
	\dot{\hat{\rho}}_{s}=\;&-i\left[\hat{H}_{\rm 2pd},\hat{\rho}_{s}\right]+\kappa_{s}\mathcal{L}\left(\hat{a}_{s}\right)\hat{\rho}_{s},\\
	\hat{H}_{\rm 2pd}=\;&\Delta_{s}\hat{a}_{s}^{\dagger}\hat{a}_{s}+\Omega_{\rm 2pd}\left(\hat{a}_{s}^{2}+{\rm H.c.}\right),
\end{align}
where $\hat{\rho}_{s}$ is the density matrix of the signal mode $\hat{a}_{s}$, $\Delta_{s}=\omega_{s}-\omega_{d}/2$, and $\Omega_{\rm 2pd}=g\alpha_{p}^{\rm d}$. Here, $\omega_{s}$ is the frequency of the signal mode, and $\omega_{d}$ is the frequency of the coherent driving of the pump mode of frequency $\omega_{p}$. Under the master equation, the number of intracavity photons, $\average{\hat{a}_{s}^{\dagger}\hat{a}_{s}}$, and the two-photon-correlation term, $\average{\hat{a}_{s}\hat{a}_{s}}$, evolve as
\begin{align}
	\average{\hat{a}_{s}^{\dagger}\hat{a}_{s}}\left(t\right)=\;&\frac{8\Omega_{\rm 2pd}^{2}}{\kappa_{s}^{2}-4\omega^{2}}-\mathcal{Q}_{0}\exp\left(-\kappa_{s}t\right),\\
	\average{\hat{a}_{s}\hat{a}_{s}}\left(t\right)=\;&-\frac{2\Omega_{\rm 2pd}\left(2\Delta_{s}+i\kappa_{s}\right)}{\kappa_{s}^{2}-4\omega^{2}}+\mathcal{Q}_{1}\exp\left(-\kappa_{s}t\right),
\end{align}
where $\omega=\sqrt{4\Omega_{\rm 2pd}^{2}-\Delta_{s}^{2}}$, and we have defined
\begin{align}
	\mathcal{Q}_{0}=\;&\frac{4\Omega_{\rm 2pd}^{2}}{\omega\left(\kappa_{s}^{2}-4\omega^{2}\right)}\left[2\omega\cosh\left(2\omega t\right)+\kappa_{s}\sinh\left(2\omega t\right)\right],\\
	\mathcal{Q}_{1}=\;&\frac{2\Omega_{\rm 2pd}}{\omega\left(\kappa_{s}^{2}-4\omega^{2}\right)}\left[\omega\left(2\Delta_{s}+i\kappa_{s}\right)\cosh\left(2\omega t\right)-\left(i2\Delta_{s}^{2}-\Delta_{s}\kappa_{s}-i8\Omega_{\rm 2pd}^{2}\right)\sinh\left(2\omega t\right)\right].
\end{align}

\begin{figure}[t]
	\centering
	\includegraphics[width=15.2cm]{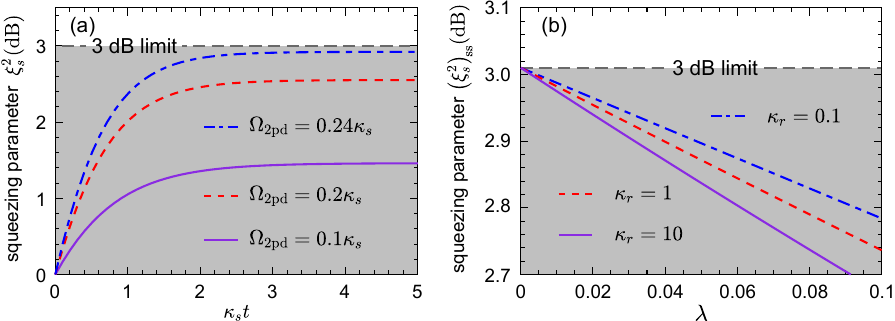}
	\caption{(a) Upper-bounded steady-state intracavity squeezing of the semiclassical degenerate parametric amplifier. Here, we set $\Delta_{s}=0$ as an example, so that the steady-state squeezing of the signal mode $\hat{a}_{s}$ approaches the $3$~dB limit as $\Omega_{\rm 2pd}\rightarrow\kappa_{s}/4$. Note that when $\Omega_{\rm 2pd}>\kappa_{s}/4$, the system becomes unstable. (b) Effects of nonlinear corrections from the residual coupling $\hat{V}$ on the steady-state intracavity squeezing, calculated according to the formalism of Ref.~\cite{chaturvedi2002limits}  for $\delta=-\lambda\sqrt{\left(2+3\kappa_{r}\right)/\left(2+\kappa_{r}\right)}/2$. As $\lambda\rightarrow0$, the steady-state intracavity squeezing increases to the 3~dB limit. In both panels, the gray shaded areas refer to the regime below the $3$~dB limit. Note that especially in panel (b), the position of the 3~dB limit is slightly greater than 3~dB, due to the fact that $-10\log_{10}(0.5)\simeq3.01$.}\label{fig_squeezing_degree_semi_classical_parametric_amplifier}
\end{figure}

To ensure that the system is stable, $\omega$ is required to be either a purely imaginary number, or a purely real number but smaller than $\kappa_{s}/2$. As a consequence, the squeezing parameter, 
\begin{equation}\label{eq_squeezing_parameter_s}
\xi_{s}^{2}=1+2\left(\average{\hat{a}_{s}^{\dagger}\hat{a}_{s}}-\left|\average{\hat{a}_{s}\hat{a}_{s}}\right|\right),
\end{equation}
in the steady state ($t\rightarrow\infty$) is found to be
\begin{equation}\label{seq_ss_sq_parameter}
	\left(\xi_{s}^{2}\right)_{\raisemath{1.2pt}{\rm ss}}=\frac{1}{1+\mu},
\end{equation}
where
\begin{equation}
	\mu=\frac{4\Omega_{\rm 2pd}}{\sqrt{\kappa_{s}^{2}-4\omega^{2}+\left(4\Omega_{\rm 2pd}\right)^{2}}}.
\end{equation}
It is clear that $\mu<1$, and thus that as shown in Fig.~\ref{fig_squeezing_degree_semi_classical_parametric_amplifier}(a), the minimum $\left(\xi_{s}^{2}\right)_{\raisemath{1.2pt}{\rm ss}}$ is limited to $0.5$, corresponding to the $3$~dB limit. 

In the above discussion, we have neglected the residual nonlinear coupling, 
\begin{equation}
\hat{V}=g\left(\hat{a}_{s}^{2}\hat{a}_{p}^{\dagger}+{\rm H.c.}\right).
\end{equation}
The effect of this residual coupling on intracavity squeezing has been considered in Ref.~\cite{chaturvedi2002limits} for $\omega_{d}=\omega_{p}=2\omega_{s}$. It was shown in Ref.~\cite{chaturvedi2002limits} that when the nonlinear corrections from the residual coupling $\hat{V}$ are taken into account, the steady-state squeezing parameter in Eq.~(\ref{seq_ss_sq_parameter}) becomes
\begin{align}\label{y_ss}
	\left(\xi_{s}^{2}\right)_{\raisemath{1.2pt}{\rm ss}}=\frac{1}{1+\mu}+\frac{\lambda^{2}\mu}{2\left(1+\mu\right)^{2}\left(1-\mu\right)}\left\{\frac{\mu\kappa_{r}}{\kappa_{r}+2}+\frac{\kappa_{r}\left(1-\mu+\mu^{2}\right)+2\left(1+\mu\right)}{\left(1+\mu\right)\left[\kappa_{r}+2\left(1+\mu\right)\right]}\right\},
\end{align}
where $\lambda=4g/\sqrt{2\kappa_{p}\kappa_{s}}$ and $\kappa_{r}=\kappa_{p}/\kappa_{s}$.
	
Because $\left(\xi_{s}^{2}\right)_{\raisemath{1.2pt}{\rm ss}}$ in Eq.~(\ref{y_ss}) becomes divergent as $\mu\rightarrow1$, we can now rewrite $\mu$ as $\mu=1+\delta$, with $\delta\lesssim0$, yielding
\begin{equation}\label{y_ss_delta}
	\left(\xi_{s}^{2}\right)_{\raisemath{1.2pt}{\rm ss}}=\frac{1}{2+\delta}-\frac{\lambda^{2}\left(1+\delta\right)}{2\delta\left(2+\delta\right)^{2}}\left[\frac{\kappa_{r}\left(1+\delta\right)}{2+\kappa_{r}}+\frac{4+\kappa_{r}+\left(2+\kappa_{r}\right)\delta+\kappa_{r}\delta^{2}}{\left(2+\delta\right)\left(4+\kappa_{r}+2\delta\right)}\right].
\end{equation}
We find, according to Eq.~(\ref{y_ss_delta}), that when
\begin{equation}
\delta=-\frac{\lambda}{2}\sqrt{\frac{2+3\kappa_{r}}{2+\kappa_{r}}}, \quad {\rm and} \quad \lambda\rightarrow0,
\end{equation}
the squeezing parameter $\left(\xi_{s}^{2}\right)_{\raisemath{1.2pt}{\rm ss}}$ reaches its minimum value $0.5$. This indicates that, as numerically confirmed in Fig.~\ref{fig_squeezing_degree_semi_classical_parametric_amplifier}(b), the steady-state intracavity squeezing is still limited to $3$~dB, even with the inclusion of the nonlinear contributions from the residual coupling $\hat{V}$.

Furthermore, we find from Fig.~\ref{fig_squeezing_degree_semi_classical_parametric_amplifier}(b) that $\left(\xi_{s}^{2}\right)_{\raisemath{1.2pt}{\rm ss}}$ decreases from the 3~dB limit, as the parameter $\lambda$, which determines the ratio of nonlinear-to-linear rates of change, increases from zero. This means that the nonlinear corrections from the residual coupling $\hat{V}$ lead to a decrease, rather than an increase, of the degree of intracavity squeezing. There are two reasons for such a decrease:
\begin{itemize}
	\item First, when the nonlinear contributions from $\hat{V}$ are considered, there exists some entanglement between the pump and signal modes, such that the squeezed state of the signal mode becomes more mixed, thus further reducing the degree of squeezing. 
	\item Second, the signal-mode photons can be up-converted into the pump mode by the residual coupling $\hat{V}$, and then can be lost via the photon loss of the pump mode. This gives rise to a two-photon loss process, which can also destroy the desired squeezing.
\end{itemize}

The 3~dB squeezing limit is essentially attributed to the cavity photon loss (i.e., the photon loss of the signal mode), which destroys the essence of squeezing, i.e., two-photon correlations [represented by the term $\average{\hat{a}_{s}\hat{a}_{s}}$ in Eq.~(\ref{eq_squeezing_parameter_s})]. This can be physically understood as follows. The two-photon driving $\Omega_{\rm 2pd}$ injects correlated photon pairs into the cavity, so theoretically generating perfect two-photon correlations. Therefore, if there is no cavity photon loss, the two-photon driving can generate an ideal, arbitrarily strong intracavity squeezing. However, the cavity photon loss is always present, such that single photons of some correlated photon pairs leak out of the cavity, thus leading to a partial loss of two-photon correlations. Consequently, intracavity squeezing is lost partially and is ultimately limited to $3$~dB in the steady state.
	
Therefore, the motivation of our method to beat the 3~dB limit is to turn the signal-mode photon loss from a noise source into a resource via quantum reservoir engineering. We find that with a fully quantum DPA, the signal-mode photon loss, when exploited as a resource, can steer the pump mode into a squeezed steady state. More importantly, we find that the signal-mode photon loss can strongly suppress the detrimental effect of the pump-mode photon loss on squeezing, and ultimately, an arbitrarily strong steady-state squeezing of the pump mode can be achieved. The underlying physical mechanism is discussed in Sec.~\ref{Physical mechanism of beating the 3dB squeezing limit}.

\section{Detailed derivation of the effective Hamiltonian}
\label{appendix_Detailed derivation of the effective Hamiltonian}
Below, we give a detailed derivation of the effective Hamiltonian $\hat{H}_{\rm eff}$. To begin, we consider the original Hamiltonian $\hat{H}=\hat{H}_{0}+\hat{H}_{\rm 2td}$, where $\hat{H}_{0}$ and $\hat{H}_{\rm 2td}$ are given in Eqs.~(1) and (2), respectively, in the main article. In terms of the signal Bogoliubov mode $\hat{\beta}_{s}$, the Hamiltonian $\hat{H}$ is reexpressed as
\begin{align}
	\label{aeq_total_Hamiltonian}
	\hat{H}=\;&\hat{\mathcal{H}}+\hat{R}_{1}+\hat{R}_{1}^{\dagger}+\hat{R}_{2}+\hat{R}_{2}^{\dagger},\\
	\label{aeq_mathcal_H}
	\hat{\mathcal{H}}=\;&\Delta_{p}\hat{a}_{p}^{\dagger}\hat{a}_{p}+\Lambda_{s}\hat{\beta}_{s}^{\dagger}\hat{\beta}_{s}+g_{0}\hat{\beta}_{s}^{\dagger}\hat{\beta}_{s}\left(\hat{a}_{p}+\hat{a}_{p}^{\dagger}\right)\nonumber\\
	&+\cosh\left(r_{s}\right)\sum_{k=\pm}\left(\mathcal{E}_{k}\hat{\beta}_{s}^{\dagger}e^{-i\omega_{k}t}+{\rm H.c.}\right),\\
	\hat{R}_{1}=\;&g\left[\cosh^{2}\left(r_{s}\right)\hat{\beta}_{s}^{2}+\sinh^{2}\left(r_{s}\right)\hat{\beta}_{s}^{\dagger2}\right]\hat{a}_{p}^{\dagger},\\
	\hat{R}_{2}=\;&-\sinh\left(r_{s}\right)\sum_{k=\pm}\mathcal{E}_{k}\hat{\beta}_{s}e^{-i\omega_{k}t}.
\end{align}
Here, $\hat{R}_{1}$ and $\hat{R}_{2}$ are the residual components of the parametric coupling $g$ and the two-tone driving $\Omega_{\rm 2td}$, respectively. Correspondingly, the master equation is given by
\begin{align}\label{aeq_full_master_equation}
	\dot{\hat{\rho}}=\;&-i\left[\hat{H},\hat{\rho}\right]+\kappa_{p}\mathcal{L}\left(\hat{a}_{p}\right)\hat{\rho}+\kappa_{s}\cosh^{2}\left(r_{s}\right)\mathcal{L}\left(\hat{\beta}_{s}\right)\hat{\rho}\nonumber\\
	&+\kappa_{s}\sinh^{2}\left(r_{s}\right)\mathcal{L}\left(\hat{\beta}_{s}^{\dagger}\right)\hat{\rho}-\frac{1}{2}\kappa_{s}\sinh\left(2r_{s}\right)\mathcal{D}\left(\hat{\beta}_{s}\right)\hat{\rho}-\frac{1}{2}\kappa_{s}\sinh\left(2r_{s}\right)\mathcal{D}\left(\hat{\beta}_{s}^{\dagger}\right)\hat{\rho},
\end{align}
where $\mathcal{L}\left(\hat{o}\right)\hat{\rho}=\hat{o}\hat{\rho} \hat{o}^{\dagger}-\frac{1}{2}\left(\hat{o}^{\dagger}\hat{o}\hat{\rho}+\hat{\rho} \hat{o}^{\dagger}\hat{o}\right)$, and $\mathcal{D}\left(\hat{o}\right)\hat{\rho}=\hat{o}\hat{\rho} \hat{o}-\frac{1}{2}\left(\hat{o}\hat{o}\hat{\rho}+\hat{\rho}\hat{o}\hat{o}\right)$.

In order to derive $\hat{H}_{\rm eff}$, let us first focus our attention on the term $\hat{\mathcal{H}}$, which provides the dominant contribution to the generation of intracavity squeezing. We then analyze the terms $\hat{R}_{1}$ and $\hat{R}_{2}$, both of which, as purely high-frequency effects, can be dropped by properly compensating some resonance shifts of the modes $\hat{\beta}_{s}$ and $\hat{a}_{p}$.
	
To proceed, we introduce a displacement operator, $\hat{D}_{p}\left(\alpha_{p}\right)=\exp\left(\alpha_{p}\hat{a}_{p}^{\dagger}-\alpha_{p}^{\dagger}\hat{a}_{p}\right)$, for the mode $\hat{a}_{p}$, and a time-dependent displacement operator, $\hat{D}_{s}\left[\alpha_{s}\left(t\right)\right]=\exp\left[\alpha_{s}\left(t\right)\hat{\beta}_{s}^{\dagger}-\alpha_{s}^{*}\left(t\right)\hat{\beta}_{s}\right]$, for the mode $\hat{\beta}_{s}$, such that
\begin{align}
\label{seq_Dp}
\hat{D}_{p}^{\dagger}\left(\alpha_{p}\right)\hat{a}_{p}\hat{D}_{p}\left(\alpha_{p}\right)=\;& \hat{a}_{p}+\alpha_{p},\\
\label{seq_Ds}
\hat{D}_{s}^{\dagger}\left[\alpha_{s}\left(t\right)\right]\hat{\beta}_{s}\hat{D}_{s}\left[\alpha_{s}\left(t\right)\right]=\;&\hat{\beta}_{s}+\alpha_{s}\left(t\right),
\end{align}
where $\alpha_{s}\left(t\right)=\alpha_{s}^{+}e^{-i\omega_{+}t}+\alpha_{s}^{-}e^{-i\omega_{-}t}$. Here, $\alpha_{p}$ is the overall field amplitude of the mode $\hat{a}_{p}$, induced by the $\omega_{\pm}$ tones (i.e., the $\omega_{\pm}^{\rm d}$ tones in the original lab frame), while $\alpha_{s}^{\pm}$ are the field amplitudes of the mode $\hat{\beta}_{s}$, induced by the $\omega_{\pm}$ tones, respectively.

When applying the displacement transformations in Eqs.~(\ref{seq_Dp}) and~(\ref{seq_Ds}), a displaced $\hat{\mathcal{H}}$ is found to be
\begin{align}
\hat{\mathcal{H}}_{\rm disp}=\;&\hat{U}^{\dagger}\left(t\right)\hat{\mathcal{H}}\hat{U}\left(t\right)-i\hat{U}^{\dagger}\left(t\right)\dot{\hat{U}}\left(t\right)\nonumber\\
&-i\frac{1}{2}\kappa_{p}\left(\alpha_{p}\hat{a}_{p}^{\dagger}-{\rm H.c.}\right)-i\frac{1}{2}\kappa_{s}\left[\alpha_{s}\left(t\right)\hat{\beta}_{s}^{\dagger}-{\rm H.c.}\right],
\end{align}
where $\hat{U}\left(t\right)=\hat{D}_{p}\left(\alpha_{p}\right)\hat{D}_{s}\left[\alpha_{s}\left(t\right)\right]$. 
It then follows, using
\begin{align}
\dot{\hat{U}}\left(t\right)=\;&\hat{D}_{p}\left(\alpha_{p}\right)\hat{D}_{s}\left[\alpha_{s}\left(t\right)\right]\nonumber\\
&\times\bigg\{\frac{1}{2}\frac{d}{dt}\left|\alpha_{s}\left(t\right)\right|^{2}-i\left(\omega_{+}\alpha_{s}^{+*}e^{i\omega_{+}t}+\omega_{-}\alpha_{s}^{-*}e^{i\omega_{-}t}\right)\left[\hat{\beta}_{s}+\alpha_{s}\left(t\right)\right]\nonumber\\
&\quad\quad-i\left(\omega_{+}\alpha_{s}^{+}e^{-i\omega_{+}t}+\omega_{-}\alpha_{s}^{-}e^{-i\omega_{-}t}\right)\hat{\beta}_{s}^{\dagger}\bigg\},
\end{align}
that 
\begin{align}\label{seq_mathcal_Hdisp2}
\hat{\mathcal{H}}_{\rm disp}
=\;&\Delta_{p}\hat{a}_{p}^{\dagger}\hat{a}_{p}+\Lambda_{s}\hat{\beta}_{s}^{\dagger}\hat{\beta}_{s}+g_{0}\hat{\beta}_{s}^{\dagger}\hat{\beta}_{s}\left(\hat{a}_{p}+\hat{a}_{p}^{\dagger}\right)
\nonumber\\
&+g_{0}\left(\alpha_{s}^{+}\hat{\beta}_{s}^{\dagger}e^{-i\omega_{+}t}+{\rm H.c.}\right)
\left(\hat{a}_{p}+\hat{a}_{p}^{\dagger}\right)+g_{0}\left(\alpha_{s}^{-}\hat{\beta}_{s}^{\dagger}e^{-i\omega_{-}t}+{\rm H.c.}\right)\left(\hat{a}_{p}+\hat{a}_{p}^{\dagger}\right)\nonumber\\
&+g_{0}\left[\alpha_{s}^{-}\left(\alpha_{s}^{+}\right)^{*}e^{i\left(\omega_{+}-\omega_{-}\right)t}+{\rm H.c.}\right]\left(\hat{a}_{p}+\hat{a}_{p}^{\dagger}\right),
\end{align}
where we have set
\begin{align}
\alpha_{p}=g_{0}\frac{\left|\alpha_{s}^{+}\right|^{2}+\left|\alpha_{s}^{-}\right|^{2}}{i\kappa_{p}/2-\Delta_{p}},\quad {\rm and} \quad \alpha_{s}^{\pm}=\frac{\cosh\left(r_{s}\right)\mathcal{E}_{\pm}}{i\kappa_{s}/2-\Lambda_{s}+\omega_{\pm}}.
\end{align} 
Note that here, the resonance shift, $2g_{0}{\rm Re}\left[\alpha_{p}\right]$, of the mode $\hat{\beta}_{s}$ has been absorbed into $\Lambda_{s}$. Under the assumption of $\omega_{\pm}=\Lambda_{s}\pm\Delta_{p}$, we can drop high-frequency components and then obtain the effective Hamiltonian in the rotating frame of reference of $\hat{h}=\Delta_{p}\hat{a}_{p}^{\dagger}\hat{a}_{p}+\Lambda_{s}\hat{\beta}_{s}^{\dagger}\hat{\beta}_{s}$ as follows: 
	\begin{align}
	\hat{H}_{\rm eff}=\;&G_{-}\hat{\beta}_{s}^{\dagger}\hat{a}_{p}+G_{+}\hat{\beta}_{s}^{\dagger}\hat{a}_{p}^{\dagger}+{\rm H.c.}\\
	\label{seq_betap0}
	=\;&\mathcal{G}\left(\hat{\beta}_{p}\hat{\beta}_{s}^{\dagger}+\hat{\beta}_{p}^{\dagger}\hat{\beta}_{s}\right),
	\end{align}
where $G_{\pm}=g_{0}\alpha_{s}^{\pm}$ and $\mathcal{G}=\sqrt{G_{-}^{2}-G_{+}^{2}}$. Here, we have set the field amplitudes $\alpha_{s}^{\pm}$ to be real, for simplicity, and have defined a Bogoliubov mode for the pump mode,
\begin{equation}\label{seq_betap}
\hat{\beta}_{p}=\hat{a}_{p}\cosh\left(r_{p}\right)+\hat{a}_{p}^{\dagger}\sinh\left(r_{p}\right),
\end{equation}
with $\tanh\left(r_{p}\right)=G_{+}/G_{-}$.
 
Now consider the term $\hat{R}_{1}$. We apply the displacement transformations in Eqs.~(\ref{seq_Dp}) and~(\ref{seq_Ds}) to $\hat{R}_{1}$, and see that $\hat{R}_{1}$ becomes
	\begin{align}\label{aeq_V10}
		\hat{R}_{1, \rm disp}=g\left[\cosh^{2}\left(r_{s}\right)\hat{\Pi}+\sinh^{2}\left(r_{s}\right)\hat{\Pi}^{\dagger}\right]\left(\hat{a}_{p}^{\dagger}+\alpha_{p}^{*}\right),
	\end{align}
	with
	\begin{align}
		\hat{\Pi}=\hat{\beta}_{s}^{2}
		+2\hat{\beta}_{s}\sum_{k=\pm}\alpha_{s}^{k}e^{-i\omega_{k}t}+\left(\sum_{k=\pm}\alpha_{s}^{k}e^{-i\omega_{k}t}\right)^{2}.
	\end{align}
As assumed in the main article, the degree, $r_{s}$, of squeezing of the signal mode $\hat{a}_{s}$ is very small, so that the terms in Eq.~(\ref{aeq_V10}), which are proportional
	to $\sinh^{2}\left(r_{s}\right)$, can be dropped, yielding
	\begin{align}\label{aeq_V11}
		\hat{R}_{1, \rm disp}\simeq\;&g_{c}\hat{\Pi}\left(\hat{a}_{p}^{\dagger}+\alpha_{p}^{*}\right)+{\rm H.c.},
	\end{align}
	where $g_{c}=g\cosh^{2}\left(r_{s}\right)\simeq g$. The coupling $\hat{R}_{1, \rm disp}$ is dominated by the couplings,
	\begin{align}
		\hat{P}=\;&2g_{c}\hat{\beta}_{s}a_{p}^{\dagger}\sum_{k=\pm}\alpha_{s}^{k}e^{-i\omega_{k}t}+{\rm H.c.},\\
		\hat{Q}=\;&g_{c}\left(\sum_{k=\pm}\alpha_{s}^{k}e^{-i\omega_{k}t}\right)^{2}\hat{a}^{\dagger}_{p}+{\rm H.c.}
	\end{align}
	Using the formalism of Ref.~\cite{gamel2010time} shows that, under the resonance conditions $\omega_{\pm}=\Lambda_{s}\pm\Delta_{p}$, the dynamics of the coupling $\hat{P}$ effectively causes a resonance up-shift of
	\begin{equation}
		\delta_{\rm shift}=2g_{c}^{2}\left[\frac{1}{\Lambda_{s}}\left(\alpha_{s}^{+}\right)^2+\frac{1}{\Lambda_{s}-\Delta_{p}}\left(\alpha_{s}^{-}\right)^2\right]
	\end{equation}
	for the mode $\hat{\beta}_{s}$, and also a resonance down-shift of the same amount for the mode $\hat{a}_{p}$. These two shifts can be compensated by slightly modifying the resonance condition $\omega_{-}=\Lambda_{s}-\Delta_{p}$ to be $\omega_{-}=\Lambda_{s}-\Delta_{p}+2\delta_{\rm shift}$.
	
	On the other hand, the coupling $\hat{Q}$ can be simply viewed as a detuned three-tone driving of the mode $\hat{a}_{p}$. To cancel its effect, a direct method is to drive the mode $\hat{a}_{p}$ using another opposite-phase three-tone driving. But to simplify this problem, we note that recently, a strong single-photon parametric coupling has been experimentally realized, with the strength $g$ being able to range from some tens of kHz to some tens of MHz~\cite{leghtas2015confining,touzard2018coherent,lescanne2020exponential,chang2020observation,vrajitoarea2020quantum,guo2016second,bruch2019chip,wang2021efficient}. This indicates that even when the intracavity-field amplitudes $\alpha_{s}^{\pm}$ are small [e.g., $\alpha_{s}^{\pm}\sim1$ as in Fig.~1(b)], a large $\mathcal{C}$ can still be achieved. Here, $\mathcal{C}=\mathcal{G}^{2}/\left(\kappa_{s}\kappa_{p}\right)$
is a key factor, whose larger value can lead to a stronger intracavity squeezing as discussed in the main article. Thus in this case, the coupling $\hat{Q}$ acts as a high-frequency component.
	At the same time, the residual driving term $\hat{R}_{2}$ can also be treated as a high-frequency component, because for $r_{s}\ll1$ we can have $\left(\omega_{\pm}+\Lambda_{s}\right)\gg \sinh\left(r_{s}\right)\mathcal{E}_{\pm}$. Consequently, both the terms $\hat{Q}$ and $\hat{R}_{2}$ are allowed to be directly dropped. Such is the case considered in the main article.

By the above analysis, it is seen that the coherent dynamics of the three-tone driven DPA, which is described exactly by the Hamiltonian $\hat{H}$, can be well governed by the effective Hamiltonian $\hat{H}_{\rm eff}$. This is numerically confirmed in Fig.~1(b) in the main article. 

\section{Physical interpretation of the effective Hamiltonian $\hat{H}_{\rm eff}$ in the laboratory frame. }
\label{Physical essence of H_sq}
\begin{figure*}[t!]
	\centering
	\includegraphics[width=17cm]{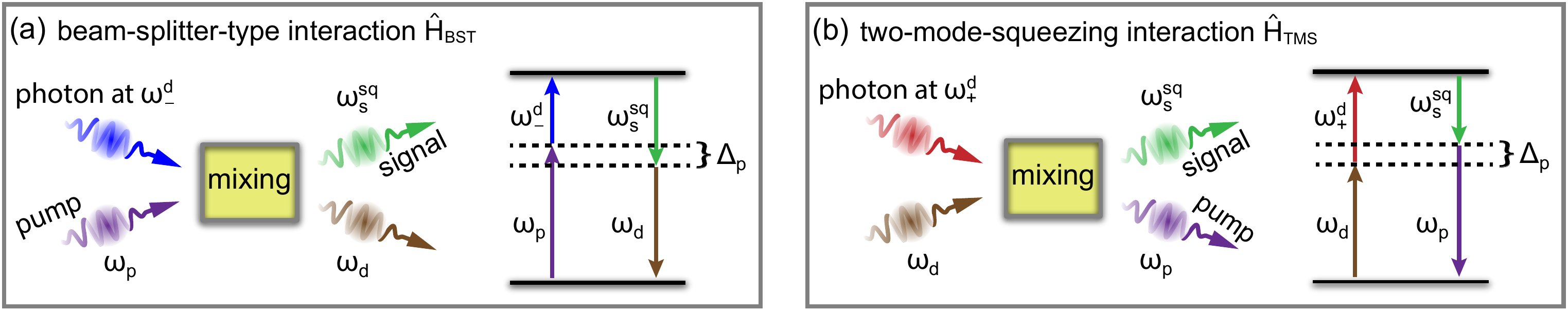}
	\caption{Four-wave-mixing processes responsible for (a) the beam-splitter-type [i.e., $\hat{H}_{\rm BST}$ in Eq.~(\ref{eq_BS_interaction})] and (b) two-mode-squeezing [i.e., $\hat{H}_{\rm TMS}$ in Eq.~(\ref{eq_TMS_interaction})] interactions in the laboratory frame. On the right of both panels is the schematic representation of energy conservation, showing the resonance conditions of these interactions, $\omega_{\pm}=\Lambda_{s}\pm\Delta_{p}$. The beam-splitter-type interaction models the conversion of a photon from the pump mode $\hat{a}_{p}$ at $\omega_{p}$ into the signal Bogoliubov mode (i.e., the shifted signal mode) $\hat{\beta}_{s}$ at $\omega_{s}^{\rm sq}$, which is accompanied by absorbing an $\omega_{-}^{\rm d}$ photon and emitting another $\omega_{d}$ photon. In contrast, the two-mode-squeezing interaction models the simultaneous creation of two photons in the signal Bogoliubov and pump modes, along with the simultaneous annihilation of an $\omega_{+}^{\rm d}$ photon and an $\omega_{d}$ photon.}\label{fig_mixing}
\end{figure*}

The effective Hamiltonian $\hat{H}_{\rm eff}$ in Eq.~(\ref{seq_betap0}) models a beam-splitter-type interaction between the signal and pump Bogoliubov modes, but {\it in the squeezed frame}. Below, we discuss how to understand this effective Hamiltonian {\it in the laboratory frame}.
To be more specific, we first note that the signal Bogoliubov mode $\hat{\beta}_{s}$ in the squeezed frame, whose resonance frequency is $\Lambda_{s}=\sqrt{\Delta_{s}^{2}-4\Omega_{\rm 2pd}^{2}}$, in fact corresponds to the shifted (or dressed) signal mode in the laboratory frame, whose resonance frequency is $\omega_{s}^{\rm sq}=\omega_{d}/2+\Lambda_{s}$. In the laboratory frame, $\hat{H}_{\rm eff}$ can be rewritten as 
\begin{align}
\hat{H}_{\rm eff}=\;& \hat{H}_{\rm BST}+\hat{H}_{\rm TMS},\\
\label{eq_BS_interaction}
\hat{H}_{\rm BST}=\;&G_{-}\left(\hat{a}_{p}\hat{\beta}_{s}^{\dagger}+\hat{a}_{p}^{\dagger}\hat{\beta}_{s}\right),\\
\label{eq_TMS_interaction}
\hat{H}_{\rm TMS}=\;&G_{+}\left(\hat{a}_{p}\hat{\beta}_{s}+\hat{a}_{p}^{\dagger}\hat{\beta}_{s}^{\dagger}\right).
\end{align}
Here, $\hat{H}_{\rm BST}$ and $\hat{H}_{\rm TMS}$ describe a resonant beam-splitter-type interaction and a resonant two-mode-squeezing interaction, respectively. Thus, the physical interpretation of $\hat{H}_{\rm eff}$ can be understood as two four-wave-mixing processes, as schematically depicted in Fig.~\ref{fig_mixing}. 

The beam-splitter-type interaction $\hat{H}_{\rm BST}$ in Eq.~(\ref{eq_BS_interaction}) exchanges a single photon between the signal Bogoliubov mode $\hat{\beta}_{s}$ at $\omega_{s}^{\rm sq}$ and the pump mode $\hat{a}_{p}$ at $\omega_{p}$. This process is obtained under the resonance condition $\omega_{-}=\Lambda_{s}-\Delta_{p}$, which, in the laboratory frame, is
\begin{align}
	\label{mixing_wave_condition_for_omega1-}
	\omega_{-}^{\rm d}+\omega_{p}=\;&\omega_{s}^{\rm sq}+\omega_{d}.
\end{align}
It is seen that the beam-splitter-type interaction $\hat{H}_{\rm BST}$ originates from a four-wave-mixing process, which is stimulated by the driving tones at $\omega_{-}^{\rm d}$ and $\omega_{d}$. According to the photon description of this process [see Fig.~\ref{fig_mixing}(a)], a pump photon at $\omega_{p}$ is down-converted to the mode $\hat{\beta}_{s}$ at $\omega_{s}^{\rm sq}$, while annihilating a driving photon at $\omega_{-}^{\rm d}$ and creating another driving photon at $\omega_{d}$.

On the other hand, the two-mode-squeezing interaction $\hat{H}_{\rm TMS}$ in Eq.~(\ref{eq_TMS_interaction}) describes the simultaneous annihilation and creation of two photons in the signal Bogoliubov mode $\hat{\beta}_{s}$ and the pump mode $\hat{a}_{p}$. Its resonance condition is $\omega_{+}=\Lambda_{s}+\Delta_{p}$, which, in the laboratory frame, reads
\begin{align}
	\label{mixing_wave_condition_for_omega1+}
	\omega_{+}^{\rm d}+\omega_{d}=\;&\omega_{s}^{\rm sq}+\omega_{p}.
\end{align}
This implies that the driving tones at $\omega_{+}^{\rm d}$ and $\omega_{d}$ simulate a four-wave-mixing process [see Fig.~\ref{fig_mixing}(b)]. In the photon description, a photon pair of the signal Bogoliubov and pump modes can be created by annihilating two driving photons at $\omega_{+}^{\rm d}$ and $\omega_{d}$.

\section{Physical mechanism of beating the 3~dB squeezing limit}
\label{Physical mechanism of beating the 3dB squeezing limit}
Having given a physical interpretation of the effective Hamiltonian $\hat{H}_{\rm eff}$ in Sec.~\ref{Physical essence of H_sq}, we here show in detail the physical mechanism of beating the 3~dB squeezing limit with our proposal. The basic idea of our proposal is to apply a two-tone driving to the signal mode, in addition to the usual driving of the pump mode. In order to show the underlying physical mechanism explicitly, we consider the effective master equation,
\begin{equation}\label{seq_effective_master_equation}
\dot{\hat{\rho}}=-i\left[\hat{H}_{\rm eff},\hat{\rho}\right]+\kappa_{p}\mathcal{L}\left(\hat{a}_{p}\right)\hat{\rho}+\kappa_{s}\mathcal{L}(\hat{\beta}_{s})\hat{\rho}.
\end{equation}
In an ideal case of $\kappa_{p}=0$, the photon loss of the mode $\hat{\beta}_{s}$ can cool the mode $\hat{\beta}_{p}$ to its ground state, i.e., the ideal squeezed vacuum state expressed as
\begin{equation}\label{seq_sqvs_Fock_basis}
\ket{\Phi_{\rm sv}}=\frac{1}{\sqrt{\cosh\left(r_{p}\right)}}\sum_{n=0}^{\infty}\frac{\sqrt{\left(2n\right)!}}{2^{n}\,n!}\left[-\tanh\left(r_{p}\right)\right]^{n}\ket{2n},
\end{equation}
in terms of photon-number states of the pump mode $\hat{a}_{p}$. According to Eq.~(\ref{seq_sqvs_Fock_basis}), the state $\ket{\Phi_{\rm sv}}$ is a superposition of only even photon-number states and, thus, has even photon-number parity. We also note that, the state $\ket{\Phi_{\rm sv}}$ is simply the zero photon-number state, $\ket{0}_{\hat{\beta}_{p}}$, of the mode $\hat{\beta}_{p}$, such that
\begin{equation}\label{seq_number_state_beta_p}
\hat{\beta}_{p}\ket{\Phi_{\rm sv}}=\hat{\beta}_{p}\ket{0}_{\hat{\beta}_{p}}=0.
\end{equation}
Furthermore, the state $\ket{\Phi_{\rm sv}}$ is a unique steady state, and can be reached from any state of the mode $\hat{a}_{p}$. The reason is that any state of the mode $\hat{a}_{p}$ can be expressed in terms of the ground and excited states of the mode $\hat{\beta}_{p}$, but of these, all the excited states (i.e., all the nonzero photon-number states of the mode $\hat{\beta}_{p}$) are depopulated by the loss of the mode $\hat{\beta}_{s}$ in the steady state, leaving only the ground state $\ket{0}_{\hat{\beta}_{p}}$, i.e., the state $\ket{\Phi_{\rm sv}}$.

We now consider the effect of the photon loss of the mode $\hat{a}_{p}$ in the case of $\kappa_{p}\neq0$. The desired steady-state squeezing is independent of the initial state of the mode $\hat{a}_{p}$, as discussed above. This enables the detrimental effect of the photon loss of the mode $\hat{a}_{p}$ on squeezing to be strongly suppressed, as long as $\kappa_{s}\gg\kappa_{p}$. Such a dissipative suppression can be understood from the quantum jump approach. The action of the quantum jump operator $\hat{a}_{p}$ on the state $\ket{\Phi_{\rm sv}}$, corresponding to a single-photon loss event, yields 
\begin{align}
\hat{a}_{p}\ket{\Phi_{\rm sv}}=-\sinh\left(r_{p}\right)\ket{1}_{\hat{\beta}_{p}},
\end{align}
where $\ket{1}_{\hat{\beta}_{p}}$ is the single photon-number state of the mode $\hat{\beta}_{p}$. It is seen that a single-photon loss event completely destroys the desired squeezing. Despite this, the strong photon loss of the mode $\hat{\beta}_{s}$ can autonomously steer the state $\ket{1}_{\hat{\beta}_{p}}$ back to the ground state $\ket{0}_{\hat{\beta}_{p}}$, i.e., the state $\ket{\Phi_{\rm sv}}$. Therefore, this strong suppression of the detrimental effect of the photon loss of the mode $\hat{a}_{p}$ on squeezing is the physical reason why our proposal can beat the 3~dB squeezing limit.

\begin{figure*}[t!]
	\centering
	\includegraphics[width=2.1cm]{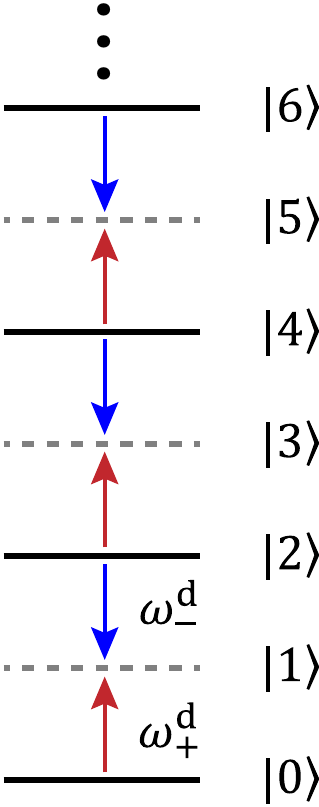}
	\caption{Schematic of destructive interference leading to an arbitrary steady-state intracavity squeezing. There always exist two different transitions to any odd photon-number state of the signal mode, which are driven by the $\omega_{\pm}^{\rm d}$ tones (red and blue arrows), respectively. A squeezed steady state, with an arbitrarily high degree of squeezing, can be achieved since these two transitions can interfere destructively, so that the populations of all the odd photon-number states are zero in the steady state.}\label{fig_destructive_interference}
\end{figure*}

Let us now discuss the role of the two driving tones, at $\omega_{\pm}^{\rm d}$, of the signal mode in our proposal. Their role is to form the Bogoliubov mode $\hat{\beta}_{p}$ in Eq.~(\ref{seq_betap}),
which involves the annihilation operator $\hat{a}_{p}$ and the creation operator $\hat{a}_{p}^{\dagger}$ simultaneously. In terms of photon-number states of the pump mode $\hat{a}_{p}$, the action of the operator $\hat{\beta}_{p}$, which is triggered by the photon loss of the mode $\hat{\beta}_{s}$, causes two different transitions to any odd photon-number state, as shown in Fig.~\ref{fig_destructive_interference}. One of them, i.e., $\ket{2n}\rightarrow\ket{2n+1}$ [corresponding to the action of the operator $\hat{a}_{p}^{\dagger}$ in Eq.~(\ref{seq_betap})], is driven by the tone at $\omega_{+}^{\rm d}$, and the other transition, i.e., $\ket{2n+2}\rightarrow\ket{2n+1}$ [corresponding to the action of the operator $\hat{a}_{p}$ in Eq.~(\ref{seq_betap})], is driven by the tone at $\omega_{-}^{\rm d}$. To satisfy Eq.~(\ref{seq_number_state_beta_p}), destructive interference between these two transitions is required to ensure that all the odd photon-number states are unpopulated in the steady state. In fact, the squeezed steady state $\ket{\Phi_{\rm sv}}$ is generated by the repeated action of the operator $\hat{\beta}_{p}$ on the pump mode, which can be initially in any state, until such destructive interference is formed in the steady state. The condition for destructive interference can also be used to derive the coefficients in Eq.~(\ref{seq_sqvs_Fock_basis}). Thus, the role of the two-tone driving is to provide two different transition pathways, which are crucial to form a strong steady-state squeezing.

\section{Longitudinal qubit readout using the semiclassical-DPA intracavity squeezing}
\label{Quantum readout of a qubit longitudinally coupled to the semiclassical DPA}
In this section, we consider the longitudinal readout of a qubit embedded in the semiclassical DPA, where the pump mode is treated as a classical field and, in turn, as a nonlinear two-photon driving of the signal mode. 
Specifically, we derive and analyze in detail the measurement signal, measurement noise, and SNR. For a semiclassical DPA, its intracavity squeezing is limited to $3$~dB, but at the same time an arbitrarily strong squeezing can, in principle, be achieved for its output field. It thus seems, at first
glance, as though the semiclassical-DPA intracavity squeezing might be suitable
to improve longitudinal qubit readout. However, below we demonstrate that {\it for the short-time measurement, the SNR does not benefit from the semiclassical-DPA intracavity squeezing, even though squeezing of the output field is very strong.} Furthermore, we also demonstrate that {\it although there is an exponential improvement in the SNR for the long-time measurement, the required measurement time is extremely long and, therefore, is infeasible experimentally.}

To begin, we assume that the qubit to be measured is longitudinally coupled to the signal mode $\hat{a}_{s}$, as described by the Hamiltonian,
\begin{align}\label{eq_longitudinal_readout_SC_Hamiltonian}
\hat{H}_{z}^{\rm sc}=\Omega_{\rm 2pd}\left(\hat{a}_{s}^{2}e^{-i2\phi_{\rm 2pd}}+\hat{a}_{s}^{\dagger2}e^{i2\phi_{\rm 2pd}}\right)+\chi_{z}\hat{\sigma}_{z}\left(\hat{a}_{s}e^{-i\phi_{z}}
+\hat{a}_{s}^{\dagger}e^{i\phi_{z}}\right),
\end{align}
where $\hat{\sigma}_{z}$ is the Pauli matrix of the qubit. The first term accounts for the semiclassical DPA with a two-photon driving of strength $\Omega_{\rm 2pd}$ and phase $2\phi_{\rm 2pd}$, and the second term for the longitudinal qubit-field coupling of strength $\chi_{z}$ and phase $\phi_{z}$. The longitudinal coupling generates a linear displacement of the signal mode $\hat{a}_{s}$, conditional on the qubit state. Under $\hat{H}_{z}^{\rm sc}$, the quantum Langevin equation of motion for the signal mode $\hat{a}_{s}$ is
\begin{align}\label{eq_longitudinal_langevin_sc}
		\dot{\hat{a}}_{s}=-i\sigma\chi_{z}e^{i\phi_{z}}-\frac{\kappa_{s}}{2}\hat{a}_{s}-i2\Omega_{\rm 2pd}\hat{a}_{s}^{\dagger}e^{i2\phi_{\rm 2pd}}-\sqrt{\kappa_{s}}\hat{a}_{s, {\rm in}}\left(t\right).
	\end{align}
	Here, the qubit has been assumed to be in a definite state, such that the operator $\hat{\sigma}_{z}$ has been written as a c-number $\sigma=\pm1$, corresponding to the excited and ground states of the qubit, respectively. This equation of motion can be solved using a formal integration, and the resulting expression for $\hat{a}_{s}\left(t\right)$ is
	\begin{align}\label{eq_as_longitudinal_sc}
		\hat{a}_{s}\left(t\right)=\;&\cosh\left[2\Omega_{\rm 2pd}\left(t-t_{0}\right)\right]\hat{a}_{s}\left(t_{0}\right)e^{-\kappa_{s}\left(t-t_{0}\right)/2}\nonumber\\
		&-ie^{i2\phi_{\rm 2pd}}\sinh\left[2\Omega_{\rm 2pd}\left(t-t_{0}\right)\right]\hat{a}_{s}^{\dagger}\left(t_{0}\right)e^{-\kappa_{s}\left(t-t_{0}\right)/2}\nonumber\\
		&-\sqrt{\kappa_{s}}\int_{t_{0}}^{t}ds\cosh\left[2\Omega_{\rm 2pd}\left(t-s\right)\right]\left[\hat{a}_{s, \rm in}\left(t\right)+ie^{i\phi_{z}}\sigma\chi_{z}/\sqrt{\kappa_{s}}\right]e^{-\kappa_{s}\left(t-s\right)/2}\nonumber\\
		&+ie^{i2\phi_{\rm 2pd}}\sqrt{\kappa_{s}}\int_{t_{0}}^{t}ds\sinh\left[2\Omega_{\rm 2pd}\left(t-s\right)\right]\left[\hat{a}_{s, \rm in}^{\dagger}\left(t\right)-ie^{-i\phi_{z}}\sigma\chi_{z}/\sqrt{\kappa_{s}}\right]e^{-\kappa_{s}\left(t-s\right)/2},
	\end{align}
	where $t_{0}$ is the initial time of the measurement.
	It is seen that in the case of $\Omega_{\rm 2pd}>\kappa_{s}/4$, the amplitude of $\hat{a}_{s}$ increases exponentially with time, and the system becomes unstable. Thus, in order for the
	system to be stable, we restrict our discussion to the case where $\Omega_{\rm 2pd}<\kappa_{s}/4$. Note that the special case of $\Omega_{\rm 2pd}=0$ corresponds to the standard longitudinal readout with no squeezing.

With the output field $\hat{a}_{s, \rm out}\left(t\right)=\hat{a}_{s, \rm in}\left(t\right)+\sqrt{\kappa_{s}}\hat{a}_{s}\left(t\right)$~\cite{PhysRevA.31.3761}, the output quadrature, carrying information about the qubit state and measured by a homodyne setup with a detection angle $\phi_{h}$, is given by
\begin{equation}\label{eq_output_Z_sc}
\hat{\mathcal{Z}}_{\rm out}\left(t\right)=\hat{a}_{s, \rm out}\left(t\right)e^{-i\phi_{h}}+\hat{a}_{s, \rm out}^{\dagger}\left(t\right)e^{i\phi_{h}}.
\end{equation}
To evaluate the SNR, which is an essential parameter quantifying the homodyne detection, we define the following measurement operator,
\begin{equation}\label{eq_measurement_operator_sc}
\hat{M}=\sqrt{\kappa_{s}}\int_{0}^{\tau}dt\,\hat{\mathcal{Z}}_{\rm out}\left(t\right),
\end{equation}
where $\tau$ is the measurement time. Its average, $\average{\hat{M}}$, is the qubit-state-dependent measurement signal,  while the variance of the noise operator $\hat{M}_{N}=\hat{M}-\average{\hat{M}}$ represents the measurement noise. Then, the SNR reads
\begin{align}\label{eq_SNR_definition}
{\rm SNR}=\frac{\left|\average{\hat{M}}_{\uparrow}-\average{\hat{M}}_{\downarrow}\right|}{\sqrt{\average{\hat{M}_{N}^{2}}_{\uparrow}+\average{\hat{M}_{N}^{2}}_{\downarrow}}},
\end{align}
where the arrows $\uparrow$ and $\downarrow$ refers to the excited and ground states of the qubit. 

Note that the longitudinal coupling in fact offers a qubit-state-dependent resonant driving for the signal mode $\hat{a}_{s}$, and can act as an internal measurement tone. In this case, the input measurement tone is no longer needed, such that the average of the input operator $\hat{a}_{s, \rm in}\left(t\right)$ equals zero, i.e., $\average{\hat{a}_{s, \rm in}\left(t\right)}=0$, and then the correlations for $\hat{a}_{s, {\rm in}}\left(t\right)$ are given by
\begin{align}
\average{\hat{a}_{s, {\rm in}}\left(t\right)\hat{a}^{\dagger}_{s, {\rm in}}\left(t^{\prime}\right)}=\;&\delta\left(t-t^{\prime}\right),\\
\average{\hat{a}_{s, {\rm in}}^{\dagger}\left(t\right)\hat{a}_{s, {\rm in}}\left(t^{\prime}\right)}=\;&\average{\hat{a}_{s, {\rm in}}\left(t\right)\hat{a}_{s, {\rm in}}\left(t^{\prime}\right)}=0.
\end{align}
Under the initial condition $\average{\hat{a}_{s}\left(0\right)}=0$, we find
\begin{align}\label{eq_longitudinal_sc_signal_separation}
\average{\hat{M}}_{\uparrow}-\average{\hat{M}}_{\downarrow}=\;&\frac{8\chi_{z}\kappa_{s}}{\left(\kappa_{s}^{2}-16\Omega_{\rm 2pd}^{2}\right)^{2}}\bigg\{\left[\kappa_{s}^{2}\left(2-\kappa_{s}\tau\right)+16\left(2+\kappa_{s}\tau\right)\Omega_{\rm 2pd}^{2}\right]\sin\left(\phi_{h}-\phi_{z}\right)\nonumber\\
&-4\Omega_{\rm 2pd}\left[\kappa_{s}\left(4-\kappa_{s}\tau\right)+16\Omega_{\rm 2pd}^{2}\tau\right]\cos\left(\phi_{h}+\phi_{z}-2\phi_{\rm 2pd}\right)\bigg\}\nonumber\\
&+\mathcal{F}_{-}e^{-\frac{1}{2}\left(\kappa_{s}-4\Omega_{\rm 2pd}\right)\tau}+\mathcal{F}_{+}e^{-\frac{1}{2}\left(\kappa_{s}+4\Omega_{\rm 2pd}\right)\tau},
\end{align}
with
\begin{equation}
\mathcal{F}_{\pm}=\frac{8\chi_{z}\kappa_{s}}{\left(\kappa_{s}\pm 4\Omega_{\rm 2pd}\right)^{2}}\left[\cos\left(\phi_{h}+\phi_{z}-2\phi_{\rm 2pd}\right)\pm\sin\left(\phi_{h}-\phi_{z}\right)\right].
\end{equation}
Here, we have assumed $t_{0}=0$ for simplicity.

Furthermore, the quantum fluctuation of the output field $\hat{a}_{s, \rm out}\left(t\right)$, given by $\hat{f}_{\rm out}\left(t\right)=\hat{a}_{s, \rm out}\left(t\right)-\average{\hat{a}_{s, \rm out}\left(t\right)}$, is found to be
\begin{align}\label{eq_fluctuation_operator_sc_longitudinal}
\hat{f}_{\rm out}\left(t\right)=\;&\hat{a}_{s, \rm in}\left(t\right)-\kappa_{s}\int_{-\infty}^{t}ds\cosh\left[2\Omega_{\rm 2pd}\left(t-s\right)\right]e^{-\kappa_{s}\left(t-s\right)/2}\hat{a}_{s, \rm in}\left(s\right)\nonumber\\
&+ie^{i2\phi_{\rm 2pd}}\kappa_{s}\int_{-\infty}^{t}ds\sinh\left[2\Omega_{\rm 2pd}\left(t-s\right)\right]e^{-\kappa_{s}\left(t-s\right)/2}\hat{a}_{s, \rm in}^{\dagger}\left(s\right).
\end{align}
Here, we have assumed that before the measurement starts (i.e., the longitudinal qubit-field coupling starts), the system (i.e., the semiclassical DPA) is already in the steady state. Thus in Eq.~(\ref{eq_fluctuation_operator_sc_longitudinal}), the lower limit of integration has been extended to $-\infty$.
The measurement noise, when expressed in terms of $\hat{f}_{\rm out}\left(t\right)$, is
\begin{align}\label{eq_expanded_measurement_operator}
\average{\hat{M}_{N}^{2}}=\kappa_{s}\int_{0}^{\tau}\int_{0}^{\tau}dt_{1}dt_{2}\bigg[&\average{\hat{f}^{\dagger}_{\rm out}\left(t_{1}\right)\hat{f}_{\rm out}\left(t_{2}\right)}+\average{\hat{f}_{\rm out}\left(t_{1}\right)\hat{f}^{\dagger}_{\rm out}\left(t_{2}\right)}\nonumber\\
&+\average{\hat{f}_{\rm out}\left(t_{1}\right)\hat{f}_{\rm out}\left(t_{2}\right)}e^{-i2\phi_{h}}+\average{\hat{f}^{\dagger}_{\rm out}\left(t_{1}\right)\hat{f}^{\dagger}_{\rm out}\left(t_{2}\right)}e^{i2\phi_{h}}\bigg],
\end{align}
and after a straightforward but tedious calculation is found as follows:
\begin{align}\label{eq_longitudinal_sc_noise}
\average{\hat{M}_{N}^{2}}=\;&\kappa_{s}\tau+\frac{16\kappa_{s}^{2}\Omega_{\rm 2pd}}{\left(\kappa_{s}^{2}-16\Omega_{\rm 2pd}^{2}\right)^{3}}\bigg\{\left[\kappa_{s}^{3}\left(2-\kappa_{s}\tau\right)+32\left(3\kappa_{s}+8\Omega_{\rm 2pd}^{2}\tau\right)\Omega_{\rm 2pd}^{2}\right]\sin\left[2\left(\phi_{h}-\phi_{\rm 2pd}\right)\right]\nonumber\\
&-8\kappa_{s}^{2}\left(3-\kappa_{s}\tau\right)\Omega_{\rm 2pd}-128\left(1+\kappa_{s}\tau\right)\Omega_{\rm 2pd}^{3}\bigg\}\nonumber\\
&+\mathcal{Z}_{-}e^{-\frac{1}{2}\left(\kappa_{s}-4\Omega_{\rm 2pd}\right)\tau}-\mathcal{Z}_{+}e^{-\frac{1}{2}\left(\kappa_{s}+4\Omega_{\rm 2pd}\right)\tau},
\end{align}
where
\begin{equation}
\mathcal{Z}_{\pm}=\frac{16\kappa_{s}^{2}\Omega_{\rm 2pd}}{\left(\kappa_{s}\pm4\Omega_{\rm 2pd}\right)^{3}}\left\{1\pm\sin\left[2\left(\phi_{h}-\phi_{\rm 2pd}\right)\right]\right\}.
\end{equation}
From Eq.~(\ref{eq_longitudinal_sc_noise}) the noise is minimized by choosing $\phi_{h}-\phi_{\rm 2pd}=\pi/4$, corresponding to the homodyne detection along the direction of squeezing. Then according to Eq.~(\ref{eq_longitudinal_sc_signal_separation}), the signal separation $\left|\average{\hat{M}}_{\uparrow}-\average{\hat{M}}_{\downarrow}\right|$ is maximized for $\phi_{h}-\phi_{z}=\pi/2$.

For these optimal phases, below we discuss the SNR. Let us first consider a special case of $\Omega_{\rm 2pd}=0$, corresponding to the standard longitudinal readout with no squeezing. In such a case, the SNR is given by~\cite{didier2015fast}
\begin{equation}\label{eq_SNR_std_longitudinal_readout}
{\rm SNR}_{z}^{\rm std}=\frac{6}{\kappa_{s}\tau}{\rm SNR}_{x}^{\rm std}=\chi_{z}\tau\sqrt{2\kappa_{s}\tau},
\end{equation}
for a short-time measurement (i.e., $\kappa_{s}\tau\ll1$). Here, ${\rm SNR}_{x}^{\rm std}=\varepsilon\left(\kappa_{s}\tau\right)^{5/2}/(3\sqrt{2}\kappa_{s})$ refers to the short-time SNR of the standard dispersive readout with a measurement tone of amplitude $\varepsilon$~\cite{wang2019ideal}. For comparison, we have assumed the amplitude $\varepsilon$ to be equal to the longitudinal coupling strength $\chi_{z}$. Clearly, the longitudinal readout requires much shorter measurement times, compared to the dispersive readout.

\begin{figure}[t]
	\centering
	\includegraphics[width=8.0cm]{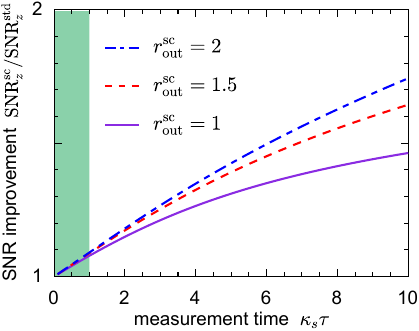}
	\caption{SNR improvement, i.e., ${\rm SNR}_{z}^{\rm sc}/{\rm SNR}_{z}^{\rm std}$, of longitudinal qubit readout using intracavity squeezing of the semiclassical degenerate parametric amplifier for $r_{\rm out}^{\rm sc}=1$, $1.5$, and $2$. The SNR is calculated using Eqs.~(\ref{eq_longitudinal_sc_signal_separation}) and~(\ref{eq_longitudinal_sc_noise}) for $\phi_{h}=\phi_{\rm 2pd}+\pi/4=\phi_{z}+\pi/2$. The green shaded area refers to the regime of most interest in experiments.}\label{fig_SNR_sc}
\end{figure}

We now consider the SNR for any $\Omega_{\rm 2pd}$ that can keep the system stable, i.e., $\Omega_{\rm 2pd}<\kappa_{s}/2$.  For the short-time measurement, i.e., $\kappa_{s}\tau\ll1$ or $\left(\kappa_{s}\pm4\Omega_{\rm 2pd}\right)\ll1$, the signal separation is found to be
\begin{equation}\label{eq_longitudianl_sc_signal_sep}
\left|\average{\hat{M}}_{\uparrow}-\average{\hat{M}}_{\downarrow}\right|\simeq2\chi_{z}\kappa_{s}\tau^{2},
\end{equation}
up to second order, and likewise, the measurement noise is reduced to
\begin{equation}\label{eq_longitudinal_sc_min_noise}
\average{\hat{M}_{N}^{2}}\simeq\kappa_{s}\tau-\frac{4\Omega_{\rm 2pd}}{\kappa_{s}+4\Omega_{\rm 2pd}}\left(\kappa_{s}\tau\right)^{2}\simeq\kappa_{s}\tau.
\end{equation}
We find that both the short-time signal and noise are almost unaffected by the two-photon driving $\Omega_{\rm 2pd}$ or equivalently by intracavity squeezing.
In turn, the optimal SNR is found to be
\begin{equation}
{\rm SNR}^{\rm sc}_{z}\simeq{\rm SNR}_{z}^{\rm std}.
\end{equation}
This implies that the semiclassical-DPA intracavity squeezing does not improve the SNR for the short-time measurement.
In Fig.~\ref{fig_SNR_sc}, we plot the SNR improvement, i.e., the ratio ${\rm SNR}_{z}^{\rm sc}/{\rm SNR}_{z}^{\rm std}$, as a function of the measurement time $\kappa_{s}\tau$. In this figure, we define the parameter
\begin{equation}
r_{\rm out}^{\rm sc}=\ln\left(\frac{\kappa_{s}+4\Omega_{\rm 2pd}}{\kappa_{s}-4\Omega_{\rm 2pd}}\right),
\end{equation}
which characterizes the degree of squeezing of the output field of the semiclassical DPA in the absence of the qubit. It can be seen that the SNR is almost unaffected in the regime $\tau\leq1/\kappa_{s}$, which is of most interest in experiments. 

Moreover, for the long-time measurement, i.e., $\kappa_{s}\tau\gg1$ or $\left(\kappa_{s}\pm4\Omega_{\rm 2pd}\right)\gg1$, an exponential improvement in the SNR can be obtained as
	\begin{equation}\label{eq_longitudinal_sc_SNR_enhancement_long_time}
		{\rm SNR}_{z}^{\rm sc}\simeq \frac{\kappa_{s}}{\kappa_{s}+4\Omega_{\rm 2pd}}\exp\left(r_{\rm out}^{\rm sc}\right){\rm SNR}_{z}^{\rm std}.
	\end{equation}
For example, choosing $r_{\rm out}^{\rm sc}=2$ ($\simeq17$~dB) can give a more than four-fold improvement in the long-time limit. However, as shown in Fig.~\ref{fig_SNR_sc}, even with a long measurement time of up to $\tau=10/\kappa_{s}$, the output-field squeezing of $r_{\rm out}^{\rm sc}=2$ can only lead to a modest improvement of the SNR by a factor of $\simeq1.7$. This means that in order to obtain the exponential improvement given in Eq.~(\ref{eq_longitudinal_sc_SNR_enhancement_long_time}), the measurement time needs to be extremely long, which is infeasible in experiments. Hence, the semiclassical-DPA intracavity squeezing {\it cannot} significantly improve the SNR during a practically feasible measurement time, even with a strong output-field squeezing. 

\begin{figure}[t]
	\centering
	\includegraphics[width=8.56cm]{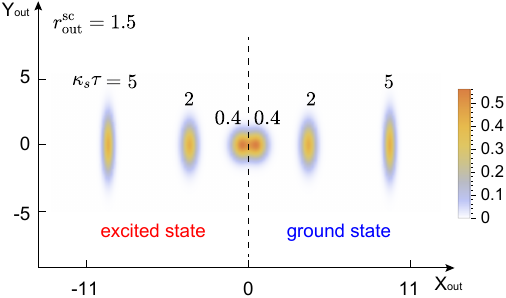}
	\caption{Phase-space representation of longitudinal qubit readout using intracavity squeezing of the semiclassical degenerate parametric amplifier. We chose, for $r_{\rm out}^{\rm sc}=1.5$ ($\simeq13$~dB), three different measurement times: $\kappa_{s}\tau=0.4$, $2$, and $5$. The resulting Wigner functions on the left and right of the vertical dashed line correspond to the excited and ground states of the qubit, respectively. The degree of squeezing of the measurement noise strongly depends on the measurement time.}\label{fig_sc_phase_space}
\end{figure}

To understand this result of no significant improvement of practical interest in experiments, we now turn to phase space. In order to describe the Wigner function of the output field, we define the temporal mode~\cite{strandberg2021wigner,lu2021propagating}
	\begin{equation}
		\hat{A}=\frac{1}{\sqrt{\tau}}\int_{0}^{\tau}dt\;\hat{a}_{s,{\rm out}}\left(t\right).
	\end{equation}
	The resulting bosonic commutation relation $\left[\hat{A},\hat{A}^{\dagger}\right]=1$ allows us to introduce two conjugate quadratures,
	\begin{align}
		\hat{X}_{\rm out}=\;\frac{1}{2}\left(\hat{A}+\hat{A}^{\dagger}\right),\quad {\rm and} \quad
		\hat{Y}_{\rm out}=\;\frac{1}{2i}\left(\hat{A}-\hat{A}^{\dagger}\right),
	\end{align}
	which satisfy $\left[\hat{X}_{\rm out},\hat{Y}_{\rm out}\right]=i$.
	With these quadratures, the Wigner function of the output field can be calculated as
	\begin{equation}\label{eq_output_wigner_function}
		W\left(X_{\rm out},Y_{\rm out}\right)=\frac{1}{2\pi\sqrt{{\rm Det}\left({\bf D}\right)}}\exp\left(-\frac{1}{2}{\bf G}^{\rm T}{\bf D}^{-1}{\bf G}\right),
	\end{equation}
	where
	\begin{align}
		{\bf G}=\;\left(X_{\rm out}-\average{\hat{X}_{\rm out}},Y_{\rm out}-\average{\hat{Y}_{\rm out}}\right)^{\rm T},
	\end{align}
	\vspace{-4mm}
	\begin{align}
		{\bf D}=\;&\left(
		\begin{array}{cc}
			\average{\hat{X}_{\rm out}^{2}}-\average{\hat{X}_{\rm out}}^{2} & \average{\hat{X}_{\rm out}\hat{Y}_{\rm out}+\hat{Y}_{\rm out}\hat{X}_{\rm out}}/2-\average{\hat{X}_{\rm out}}\average{\hat{Y}_{\rm out}} \\
			\average{\hat{X}_{\rm out}\hat{Y}_{\rm out}+\hat{Y}_{\rm out}\hat{X}_{\rm out}}/2-\average{\hat{X}_{\rm out}}\average{\hat{Y}_{\rm out}}  &  \average{\hat{Y}_{\rm out}^{2}}-\average{\hat{Y}_{\rm out}}^{2} \\
		\end{array}
		\right).
	\end{align}
	We show in Fig.~\ref{fig_sc_phase_space} the Wigner functions in phase space for different measurement times and for the ground and excited states of the qubit. The degree of squeezing of the measurement noise increases from the initial value zero as the measurement time increases, and then tends asymptotically towards the limit value $r_{\rm out}^{\rm sc}$. The practically unsqueezed measurement noise in the regime  $\kappa_{s}\tau\leq1$ is the physical reason for almost no improvement in the SNR  for the short-time measurement.

\section{Improved longitudinal qubit readout using our fully-quantum-DPA intracavity squeezing}
\label{Enhanced longitudinal qubit readout using our approach}
Following the same method as in Sec.~\ref{Quantum readout of a qubit longitudinally coupled to the semiclassical DPA}, we here derive and analyze the measurement signal, measurement noise, and SNR for the longitudinal readout of a qubit embedded in the fully quantum DPA. In our approach, the qubit is coupled to the pump mode, rather than to the signal mode as described in Sec.~\ref{Quantum readout of a qubit longitudinally coupled to the semiclassical DPA}. We demonstrate that, compared to the case of using the semiclassical-DPA intracavity squeezing, {\it our fully-quantum-DPA intracavity squeezing enables an exponential improvement in the SNR for any measurement time, thus yielding a faster and higher-SNR (i.e., higher-fidelity) qubit readout}.
	
With the Hamiltonian $\hat{H}_{z}^{\rm fq}$ in Eq.~(14) in the main article,
the quantum Langevin equations of motion for the modes $\hat{a}_{p}$ and $\hat{\beta}_{s}$ are
	\begin{align}
		\dot{\hat{a}}_{p}=\;&-i\sigma\chi_{z}e^{i\phi_{z}}-i\left(G_{-}\hat{\beta}_{s}+G_{+}\hat{\beta}_{s}^{\dagger}\right)-\frac{\kappa_{p}}{2}\hat{a}_{p}-\sqrt{\kappa_{p}}\hat{a}_{p, \rm in}\left(t\right),\\
		\dot{\hat{\beta}}_{s}=\;&-i\left(G_{-}\hat{a}_{p}+G_{+}\hat{a}_{p}^{\dagger}\right)-\frac{\kappa_{s}}{2}\hat{\beta}_{s}-\sqrt{\kappa_{s}}\hat{\beta}_{s, \rm in}.
	\end{align}
	As was assumed for the generation of the strong steady-state intracavity squeezing, the photon loss rate $\kappa_{s}$ of the mode $\hat{\beta}_{s}$ is sufficiently large, such that we can set $\dot{\hat{\beta}}_s=0$ to adiabatically eliminate the mode $\hat{\beta}_{s}$. This leads to the following adiabatic equation of motion for the mode $\hat{a}_{p}$,
	\begin{equation}
		\dot{\hat{a}}_{p}=-i\sigma\chi_{z}e^{i\phi_{z}}-\frac{\kappa}{2}\hat{a}_{p}-\sqrt{\kappa}\hat{\mathcal{A}}_{\rm in}\left(t\right).
	\end{equation}
	Here, the noise operator $\hat{\mathcal{A}}_{\rm in}\left(t\right)=\frac{1}{\sqrt{\kappa}}\left[\sqrt{\kappa_{p}^{\rm ad}}\hat{a}_{p,{\rm in}}^{\rm ad}\left(t\right)+\sqrt{\kappa_{p}}\hat{a}_{p,{\rm in}}\left(t\right)\right]$ describes the overall input noise. The pump mode $\hat{a}_{p}$ is then found, after formally integrating, to be
	\begin{align}
		\hat{a}_{p}\left(t\right)=\;&e^{-\kappa\left(t-t_{0}\right)/2}\hat{a}_{p}\left(t_{0}\right)\nonumber\\
		&-i\frac{2}{\kappa}\sigma\chi_{z}e^{i\phi_{z}}\left[1-e^{-\kappa\left(t-t_{0}\right)/2}\right]-\sqrt{\kappa}\int_{t_{0}}^{t}e^{-\kappa\left(t-s\right)/2}\hat{\mathcal{A}}_{\rm in}\left(s\right)ds.
	\end{align}
	
	As mentioned in Sec.~\ref{Quantum readout of a qubit longitudinally coupled to the semiclassical DPA}, the longitudinal coupling can be thought of as an internal measurement tone, and there is no input or external measurement tone. In this case, $\average{\hat{\beta}_{s,{\rm in}}\left(t\right)}=\average{\hat{\mathcal{A}}_{\rm in}\left(t\right)}=0$. Moreover, the correlations for $\hat{\beta}_{s,{\rm in}}\left(t\right)$, which approximately describe the vacuum noise of the mode $\hat{\beta}_{s}$, are
	\begin{align}
		\average{\hat{\beta}_{s,{\rm in}}\left(t\right)\hat{\beta}_{s,{\rm in}}^{\dagger}\left(t^{\prime}\right)}=\;&\delta\left(t-t^{\prime}\right),\\
		\average{\hat{\beta}_{s,{\rm in}}^{\dagger}\left(t\right)\hat{\beta}_{s,{\rm in}}\left(t^{\prime}\right)}=\;&\average{\hat{\beta}_{s,{\rm in}}\left(t\right)\hat{\beta}_{s,{\rm in}}\left(t^{\prime}\right)}=0,
	\end{align}
	and, as a consequence, the correlations for $\hat{\mathcal{A}}_{\rm in}\left(t\right)$ are
	\begin{align}
		\label{eq_correlation_1}
		\average{\hat{\mathcal{A}}_{\rm in}\left(t\right)\hat{\mathcal{A}}_{\rm in}^{\dagger}\left(t^{\prime}\right)}=\;&\left[\frac{1}{4\mathcal{C}+1}+\frac{4\mathcal{C}}{4\mathcal{C}+1}\cosh^{2}\left(r_{p}\right)\right]\delta\!\left(t-t^{\prime}\right),\\
		\average{\hat{\mathcal{A}}_{\rm in}^{\dagger}\left(t\right)\hat{\mathcal{A}}_{\rm in}\left(t^{\prime}\right)}=\;&\frac{4\mathcal{C}}{4\mathcal{C}+1}\sinh^{2}\left(r_{p}\right)\delta\!\left(t-t^{\prime}\right),\\
		\average{\hat{\mathcal{A}}_{\rm in}\left(t\right)\hat{\mathcal{A}}_{\rm in}\left(t^{\prime}\right)}=\;&-\frac{2\mathcal{C}}{4\mathcal{C}+1}\sinh\left(2r_{p}\right)\delta\!\left(t-t^{\prime}\right).
	\end{align}
	Here, $\mathcal{C}=\mathcal{G}^{2}/\left(\kappa_{s}\kappa_{p}\right)$ is the cooperativity of the DPA. It is seen that the correlation in Eq.~(\ref{eq_correlation_1}) includes two contributions, one from the natural photon loss of the pump mode $\hat{a}_{p}$ and the other from the adiabatic effect of the signal Bogoliubov mode $\hat{\beta}_{s}$.
	
	It follows, according to the input-output relation $\hat{\mathcal{A}}_{\rm out}\left(t\right)=\hat{\mathcal{A}}_{\rm in}\left(t\right)+\sqrt{\kappa}\hat{a}_{p}\left(t\right)$, that the signal separation is found to be
	\begin{equation}
		\left|\average{\hat{M}}_{\uparrow}-\average{\hat{M}}_{\downarrow}\right|=8\chi_{z}\tau\left|\sin\left(\phi_{h}-\phi_{z}\right)\right|\left\{1-\frac{2}{\kappa\tau}\left[1-\exp\left(-\kappa\tau/2\right)\right]\right\}.
	\end{equation}
	Here, we have assumed $t_{0}=0$ as usual. This signal separation is the same as obtained in the standard longitudinal readout with no squeezing~\cite{didier2015fast,touzard2019gated}. Moreover, the quantum fluctuation noise of the output field now is $\hat{f}_{\rm out}\left(t\right)=\hat{\mathcal{A}}_{\rm out}\left(t\right)-\average{\hat{\mathcal{A}}_{\rm out}\left(t\right)}$, and evolves as
	\begin{equation}
		\hat{f}_{\rm out}\left(t\right)=\hat{\mathcal{A}}_{\rm in}\left(t\right)-\kappa\int_{-\infty}^{t}\exp\left[-\kappa\left(t-s\right)/2\right]\hat{\mathcal{A}}_{\rm in}\left(s\right)ds.
	\end{equation}
	Here, the lower limit of integration has been extended to $-\infty$ for the same reason as mentioned in Sec.~\ref{Quantum readout of a qubit longitudinally coupled to the semiclassical DPA}. The expression of the measurement noise is the same as in Eq.~(\ref{eq_expanded_measurement_operator}) but now replacing $\kappa_{s}\mapsto\kappa$, and, then, a straightforward calculation leads to
	\begin{equation}\label{eq_noise_longitudinal_fq}
		\average{\hat{M}_{N}^{2}}=\kappa\tau\left\{\frac{1}{4\mathcal{C}+1}+\frac{4\mathcal{C}}{4\mathcal{C}+1}\left[\cosh\left(2r_{p}\right)-\cos\left(2\phi_{h}\right)\sinh\left(2r_{p}\right)\right]\right\}.
	\end{equation}
	As long as $\mathcal{C}\gg e^{2r_{p}}/4$, then $\average{\hat{M}_{N}^{2}}$ is reduced to the measurement noise of the longitudinal readout with injecting a squeezed reservoir in the ideal case of no transmission and injection losses~\cite{didier2015fast}, i.e., 
	\begin{equation}
		\average{\hat{M}_{N}^{2}}_{\rm ideal}=\kappa\tau\left[\cosh\left(2r_{p}\right)-\cos\left(2\phi_{h}\right)\sinh\left(2r_{p}\right)\right].
	\end{equation}
	
\begin{figure}[t]
	\centering
	\includegraphics[width=8.56cm]{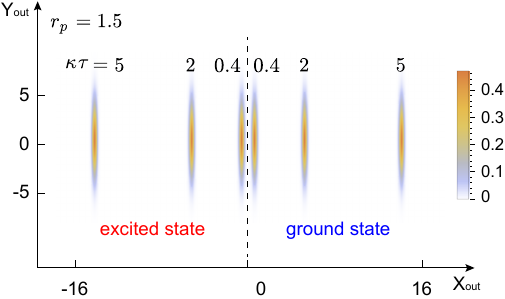}
	\caption{Phase-space representation of longitudinal qubit readout using intracavity squeezing of the fully quantum degenerate parametric amplifier. We chose, for $r_{\rm out}=1.5$ ($\simeq13$~dB), three different measurement times: $\kappa\tau=0.4$, $2$, and $5$. The resulting Wigner functions on the left and right of the vertical dashed line correspond to the excited and ground states of the qubit, respectively. The strong squeezing of the measurement noise is independent of the measurement time.}\label{fig_SNR_enhancement_fq}
\end{figure}

	By choosing $\phi_{z}=\pi/2$ and $\phi_{h}=0$, the optimal SNR of the longitudinal readout using the fully-quantum-DPA intracavity squeezing is given by
	\begin{equation}\label{eq_SNR_improvement_fq}
		{\rm SNR}_{z}^{\rm fq}=\sqrt{\frac{4\mathcal{C}+1}{4\mathcal{C}\exp\left(-2r_{p}\right)+1}}{\rm SNR}_{z}^{\rm std},
	\end{equation}
which shows a significant improvement in the SNR. In particular, an exponential improvement, \begin{align}\label{seq_exponential_enhancement}
{\rm SNR}_{z}^{\rm fq}\simeq\exp\left(r_{p}\right){\rm SNR}_{z}^{\rm std},
\end{align}
can be achieved for $\mathcal{C}\gg e^{2r_{p}}/4$. We note that Eqs.~(\ref{eq_SNR_improvement_fq}) and~(\ref{seq_exponential_enhancement}) hold for {\it any} measurement time. This is because the degree of squeezing of the measurement noise, i.e., $\average{\hat{M}_{N}^{2}}/\kappa\tau$, is independent of the measurement time $\tau$ [see Eq.~(\ref{eq_noise_longitudinal_fq})], and particularly is equal to the degree of intracavity squeezing for $\phi_{h}=0$, i.e., $\average{\hat{M}_{N}^{2}}/\kappa\tau=\left(\xi_{p}^{2}\right)_{\raisemath{1.2pt}{\rm ss}}=\left[1+4\mathcal{C}\exp\left(-2r_{p}\right)\right]/\left(1+4\mathcal{C}\right)$. This becomes more apparent in phase space shown in Fig.~\ref{fig_SNR_enhancement_fq}. We find that a short measurement time of $\tau=0.4/\kappa$ can well resolve the measurement signals associated with the ground and excited states of the qubit. This is in stark contrast to what we have already shown in Fig.~\ref{fig_sc_phase_space}, where the same measurement time (i.e., $\tau=0.4/\kappa_{s}$) leads to a high degree of overlap of the measurement signals, such that they cannot be resolved. In order to compare, we here have assumed $\kappa_{s}$ in Fig.~\ref{fig_sc_phase_space} and $\kappa$ in Fig.~\ref{fig_SNR_enhancement_fq} to be equal. Thus, our approach can enable a faster and higher-SNR qubit readout.

\section{Possible implementations of our approach for improving longitudinal qubit readout}
\label{Possible implementations of our approach for enhancing longitudinal qubit readout}
The longitudinal qubit-field coupling can be directly realized via circuit design in circuit quantum electrodynamics~\cite{liu2006controllable,kerman2013quantum,billangeon2015circuit,billangeon2015scalable,richer2016circuit,richer2017inductively,stassi2018long}. Alternatively, some synthetic approaches, e.g., strongly driving the qubit-field dispersive coupling, have also been proposed and even demonstrated in experiments~\cite{you2003quantum,blais2007quantum,eddins2018stroboscopic,touzard2019gated,ikonen2019qubit,dassonneville2020fast}. In this section, as an
example, we discuss in detail a possible implementation of the Hamiltonian $\hat{H}_{z}^{\rm fq}$ in Eq.~(14) in a synthetic manner. In particular, we refer to the
experimental implementation reported in Ref.~\cite{ikonen2019qubit}, where the longitudinal coupling is synthesized with driving the qubit at the cavity frequency.

Let us now assume that
the qubit is driven with phase $\phi_{z}$, amplitude $\mathcal{E}_{q}^{\rm d}$, and frequency $\omega_{q}^{\rm d}$. The total Hamiltonian accordingly is given, in the frame rotating at $\omega_{d}$, by
\begin{align}
	\hat{H}_{T}=\;&\hat{H}+\frac{1}{2}\Delta_{q}\hat{\sigma}_{z}+g_{q}\left(\hat{\sigma}_{-}\hat{a}_{p}^{\dagger}
	+{\rm H.c.}\right)\nonumber\\
	&+\mathcal{E}_{q}^{\rm d}\left[e^{i\phi_{z}}\hat{\sigma}_{-}e^{-i\left(\omega_{d}-\omega_{q}^{\rm d}\right)t}+{\rm H.c.}\right],
\end{align}
where $\hat{H}$ has been given in the main article, $\hat{\sigma}_{\pm}$ are the raising/lowering operators of the qubit, $g_{q}$ is the strength of the coupling of the qubit to the pump mode, and $\Delta_{q}=\omega_{q}-\omega_{d}$ is the detuning of the qubit from the driving of the pump mode.
We follow the same procedure listed in Sec.~\ref{appendix_Detailed derivation of the effective Hamiltonian} to find that the dynamics of the total system can be described by the following Hamiltonian
\begin{align}\label{eq_H_T_prime0}
	\hat{H}_{T}^{\prime}=\;&\hat{H}_{\rm eff}+g_{q}\left(\hat{\sigma}_{-}\hat{a}_{p}^{\dagger}e^{-i\Delta_{q}^{\rm d}t}+{\rm H.c.}\right)\nonumber\\
	&+\mathcal{E}_{q}^{\rm d}\left(\hat{\sigma}_{-}e^{i\phi_{z}}e^{-i\Delta_{q}^{\rm d}t}+{\rm H.c.}\right)\nonumber\\
	&+g_{q}\alpha_{p}^{\rm d}\left(\hat{\sigma}_{-}e^{-i\Delta_{q}t}+{\rm H.c.}\right).
\end{align}
Here, we have assumed that $\omega_{q}-\omega_{p}=\omega_{q}-\omega_{q}^{\rm d}=\Delta_{q}^{\rm d}$.
Note that the detuned driving of the qubit, which is associated with the field amplitudes $\alpha_{p}$ induced by the $\omega_{\pm}^{\rm d}$ drivings of the signal mode, have been neglected since $\alpha_{p}\ll\alpha_{p}^{\rm d}$.

\begin{figure}[t]
	\centering
	\includegraphics[width=8.0cm]{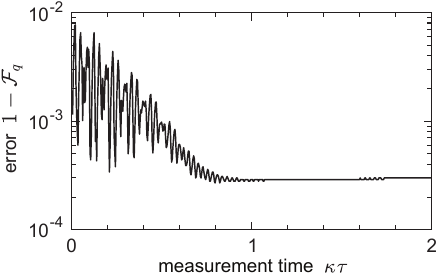}
	\caption{State error, $1-\mathcal{F}_{q}$, as a function of the measurement time $\kappa\tau$. We assumed that $g_{q}=g$, $\mathcal{E}_{q}^{\rm d}=10g_{q}$, $\phi_{z}=\pi/2$, $\Delta_{q}^{\rm d}=200g_{q}$, $\Delta_{q}=\Delta_{q}^{\rm d}+\Delta_{p}$, $G_{+}=0.7G_{-}$, and other parameters associated with $\hat{H}_{\rm eff}$ are the same as those in Fig.~1(b). We assumed that the qubit is initially in the state $\left(\ket{g}+\ket{e}\right)/\sqrt{2}$, while the signal and pump modes are in the steady state given by $\hat{H}_{\rm eff}$.}\label{fig_error}
\end{figure}

We further assume $\Delta_{q}^{\rm d}\gg \left\{g_{q}, \mathcal{E}_{q}^{\rm d}\right\}$ and $\Delta_{q}\gg g_{q}\alpha_{p}^{\rm d}$, justifying for a perturbative treatment of the second, third and fourth terms of $\hat{H}_{T}^{\prime}$ using the formalism of Ref.~\cite{gamel2010time}. We then find that
\begin{align}\label{eq_H_T_prime}
	\hat{H}_{T}^{\prime}\approx\hat{H}_{\rm eff}+\chi_{z}\hat{\sigma}_{z}\left(\hat{a}_{p}e^{-i\phi_{z}}
	+\hat{a}^{\dagger}_{p}e^{i\phi_{z}}\right)+\chi_{x}\hat{a}_{p}^{\dagger}\hat{a}_{p}\hat{\sigma}_{z},
\end{align}
where
\begin{equation}
	\chi_{z}=\mathcal{E}_{q}^{\rm d}g_{q}/\Delta_{q}^{\rm d},\quad {\rm and} \quad \chi_{x}=g^{2}_{q}/\Delta_{q}^{\rm d}
\end{equation}
are the strengths of the longitudinal and dispersive couplings between the qubit and the pump mode, respectively. In Eq.~(\ref{eq_H_T_prime}), we have subtracted the term describing the resonance shift of the qubit, i.e., $\frac{1}{2}\delta_{z}\hat{\sigma}_{z}$, where
\begin{equation}
	\delta_{z}=\left[g_{q}^{2}+2\left(\mathcal{E}_{q}^{\rm d}\right)^{2}\right]/\Delta_{q}^{\rm d}+2\left(g_{q}\alpha_{p}^{\rm d}\right)^{2}/\Delta_{q}.
\end{equation}
This is because this term can be completely eliminated in a proper frame. Under the assumption of $\mathcal{E}_{p}^{\rm d}\gg g_{q}$, the longitudinal coupling $\chi_{z}$ is much stronger than the dispersive coupling $\chi_{x}$, so that the latter can be neglected, yielding
\begin{equation}\label{eq_synthetic_longitudinal_coupling}
	\hat{H}_{T}^{\prime}\simeq\hat{H}_{\rm eff}+\chi_{z}\hat{\sigma}_{z}\left(\hat{a}_{p}e^{-i\phi_{z}}
	+\hat{a}^{\dagger}_{p}e^{i\phi_{z}}\right).
\end{equation}
This is the Hamiltonian $\hat{H}_{z}^{\rm fq}$ in Eq.~(14) in the main article.

To confirm the validity of Eq.~(\ref{eq_synthetic_longitudinal_coupling}), we perform numerical simulations, as shown in Fig.~\ref{fig_error}. Specifically, we calculate the fidelity, $\mathcal{F}_{q}$, of two qubit states that are given by the evolution under the two Hamiltonians in Eqs.~(\ref{eq_H_T_prime0}) and~(\ref{eq_synthetic_longitudinal_coupling}), respectively, from the same initial state. In Fig.~\ref{fig_error}, we assume that the initial state of the qubit is $\left(\ket{g}+\ket{e}\right)/\sqrt{2}$. At the same time, the initial state of the signal and pump modes is assumed to be the steady state given by $\hat{H}_{\rm eff}$, indicating that when the measurement starts, a steady-state squeezing has already been generated.  It can be seen in Fig.~\ref{fig_error} that during a long measurement time, the state error, $1-\mathcal{F}_{q}$, can be kept well below $10^{-2}$ for some modest parameters.

\newpage

\end{CJK*}


%